\documentclass[a4paper,UKenglish,cleveref, autoref, thm-restate]{lipics-v2021}
\usepackage[title]{appendix}
\usepackage{placeins}
\usepackage{booktabs}
\usepackage{makecell}
\newcommand{\set}[1]{\left\{{#1}\right\}}
\newcommand{\ceil}[1]{\lceil{#1}\rceil}
\newcommand{\floor}[1]{\lfloor{#1}\rfloor}
\usepackage{changepage}
    
\usepackage[colorinlistoftodos]{todonotes}
\usepackage[normalem]{ulem}

\newcommand{\bestthresholdpoint}[0]{$10^{67}$}

\usepackage{tikz}
\usepackage{xcolor}
\definecolor{colorDijkstra}{HTML}{377EB8}
\definecolor{colorBMSSP}{HTML}{E41A1C}

\pdfoutput=1 
\hideLIPIcs  

\usepackage{natbib}
\title{Implementation and Experimental Analysis of the Duan et al. (2025) Algorithm for Single-Source Shortest Paths}
\titlerunning{Experimental Analysis of the Duan et al. (2025) Algorithm for SSSP}

\author{Lucas Castro}{Institute of Computing -- UFAM, Brazil}{lucas.castro@icomp.ufam.edu.br}{https://orcid.org/0009-0008-0876-823X}{}

\author{Thailsson Clementino}{Institute of Computing -- UFAM, Brazil}{thailsson.clementino@icomp.ufam.edu.br}{https://orcid.org/0009-0009-1586-9204}{}

\author{Rosiane de Freitas}{Institute of Computing -- UFAM, Brazil}{rosiane@icomp.ufam.edu.br}{https://orcid.org/0000-0002-7608-2052}{}

\authorrunning{L. Castro, T. Clementino and R. de Freitas} 

\Copyright{Lucas Castro de Souza, Thailsson Clementino de Andrade and Rosiane de Freitas Rodrigues} 



\category{} 

\acknowledgements{
    This research was partially supported by the Coordination for the Improvement of Higher Education Personnel -- Brazil (CAPES-PROEX) -- Funding Code 001, the National Council for Scientific and Technological Development (CNPq), and the Amazonas State Research Support Foundation -- FAPEAM -- through the POSGRAD 2024-2025 project.
}

\nolinenumbers 

\begin{document}

\maketitle

\begin{abstract}

We present an implementation and experimental analysis of the deterministic algorithm proposed by Duan et al. (2025) for the Single-Source Shortest Path (SSSP) problem, which achieves the best-known asymptotic upper bound of $O(m \log^{2/3} n)$. We provide a worst-case C\texttt{++} implementation of this algorithm utilizing $O(n \log^{1/3} n)$ space, as well as a variant that reduces memory usage to $O(n)$ while maintaining the same time complexity in expectation. We compare these implementations against Dijkstra’s algorithm on sparse random graphs, grids, and U.S. road networks with up to 10 million vertices. Our results show that while the implementations adhere to their theoretical complexity bounds, large constant factors hinder their practical utility; Dijkstra's algorithm remains 3 to 4 times faster in all tested scenarios. Furthermore, we estimate that the number of vertices would need to vastly exceed \bestthresholdpoint{} for the worst-case implementation to outperform Dijkstra's. These findings suggest that a substantial reduction in constant factors is required before this theoretical breakthrough can displace established methods in practical applications.

\end{abstract}

\section{Introduction}

Let $G = (V, E)$, $|V| = n, |E| = m$, be a connected graph with a weight function 
$w: E \rightarrow \mathbb{R}_{\geq 0}$ and a source vertex $s \in V$.
The \textbf{Single-Source Shortest Path} (SSSP) problem
consists of determining, for each vertex $v \in V$, the minimum distance 
$d(v) = d(s,v)$ of a path in $G$ that starts at $s$ and ends at $v$.

As one of the most classical algorithmic problems in graph theory, 
the SSSP is widely studied in the literature. Until the end of the last century, the algorithm proposed by \citet{dijkstra1959note} was still considered the state of the art for solving the SSSP. When combined with an efficient priority queue, such as a binary heap \citep{williams_algorithm_1964}, Dijkstra's algorithm solves the problem with a time complexity of $O((n + m) \log n)$. Recently, \citet{haeupler2024universal} showed that Dijkstra's algorithm 
is universally optimal for the natural problem of ordering the vertices according to 
their distances $d(v)$ from the source $s$. In sparse graphs, where $m = O(n)$, 
this complexity matches the sorting barrier in the comparison-addition model.

To surpass this barrier, researchers have explored alternative computation models or focused on more restricted variants of the SSSP, such as those defined on particular graph types or under specific edge-weight constraints. In the RAM computation model, \citet{thorup1999undirected} presented an algorithm with time complexity 
$O(m)$ for undirected graphs whose weights fit into a machine word, 
and later \citet{thorup2004integer} extended the approach to directed graphs, 
achieving $O(m + n \log \log n)$. In the comparison-addition model, 
\citet{duan2023randomized} proposed a randomized algorithm for undirected graphs with time
complexity $O(m \sqrt{\log n \log \log n})$.

Recently, \citet{duan2025breaking} presented the first deterministic algorithm 
with $o(n \log n)$ complexity for sparse graphs, achieving an upper bound of $O(m \log^{2/3} n)$.  While this result is a landmark in theoretical computer science, its practical utility remains an open question.

Initial steps toward its implementation have recently emerged. \citet{valko2025outperforming} provided a Rust implementation evaluated on graphs of up to approximately 15k nodes. They adopted the practical approach of using sorting as their selection algorithm, but while effective, this choice does not preserve the theoretical worst-case bounds. Concurrently with our work, \citet{makowski2025bmsspy} developed a faithful Python implementation and evaluated it on grid and geospatial networks, though the choice of language limits performance analysis.

Several works, including \citet{cherkassky1996shortest}, \citet{seth2002}, \citet{goldberg2005computing}, and \citet{demetrescu2006experimental}, have conducted extensive experimental studies on shortest path problems, highlighting the importance of empirical evaluation alongside theoretical advances.

In this work, we aim to evaluate the practical behavior of the algorithm proposed by \citet{duan2025breaking} across diverse graph topologies while strictly adhering to its theoretical complexity bounds. Our contributions are as follows:
\begin{itemize}
    \item We provide a \texttt{C++} implementation that strictly follows the theoretical proposal of \citet{duan2025breaking}, utilizing $O(n \log^{1/3} n)$ space.
    \item We introduce a variant that achieves the same time complexity in expectation while utilizing only $O(n)$ memory.
    \item We assess the practical overhead of transforming the graph to constant degree, a necessary step to guarantee the theoretical complexity proposed in the paper.
    \item We conduct a comprehensive experimental evaluation on random graphs, grid graphs, and U.S. road networks, scaling up to instances with over 10 million vertices.
\end{itemize}

The remainder of this paper is organized as follows.
\Cref{sec:desc_algo} outlines the main ideas of the new algorithm.
\Cref{sec:constants} analyzes the expected asymptotic behavior of the
\citet{duan2025breaking} algorithm in comparison with
Dijkstra's algorithm.
\Cref{sec:implementation_details} details each of our implementation variants. \Cref{sec:experiments} describes the experimental setup, while \Cref{sec:results1,sec:results2,sec:results3} present the results. Finally, \Cref{sec:general} provides a general analysis of the results,
and \Cref{sec:conclusion} concludes the paper.

\section{Duan et al. (2025)'s Algorithm} \label{sec:desc_algo}

The algorithm proposed by \citet{duan2025breaking} addresses an extension of the SSSP called 
\textit{Bounded Multi-source Shortest Path} (BMSSP). In BMSSP, the goal is to compute, for each vertex $v \in V$, 
the minimum distance $d(v)$ from some vertex $u \in S \subseteq V$, subject to the constraint $d(v) < B$. 
When $S = \{s\}$ and $B = \infty$, BMSSP reduces to the standard SSSP.

The algorithm uses two parameters, $k$ and $t$, and is based on a divide-and-conquer approach over the vertex set. 
The vertex set is recursively divided into $2^t$ roughly equal parts across $O((\log n)/t)$ recursion levels, 
until reaching the base case where a subproblem contains a single vertex $x$. 
In the base case, Dijkstra's algorithm is executed to compute the shortest distances from $x$ to its $k$ nearest vertices.

The algorithm faces two main bottlenecks: 
\begin{enumerate}
    \item selecting the new set $S$ to connect sequential subproblems in a partition, and
    \item controlling the size of $S$ to prevent degradation of the overall complexity.
\end{enumerate}

To address the first bottleneck, \citet{duan2025breaking} use a specialized data structure to select the 
$2^{(l-1)t}$ vertices with the smallest distances in each partition at recursion level $l$. (This data structure has been fully implemented in this work.) To address the second bottleneck, the set $S$ is reduced to a smaller set of \emph{pivots} $P \subseteq S$ 
that suffice as sources. This reduction is performed through $k$ iterations of a Bellman-Ford-like algorithm \citep{bellman1958routing}.

The parameters $k$ and $t$ are chosen to balance the cost of the Bellman-Ford iterations with the recursion depth, 
resulting in a final time complexity of $O(m \log^{2/3} n)$.

\section{Asymptotic Versus Empirical Analysis Under Big Constants} \label{sec:constants}

Asymptotic complexity characterizes an algorithm's growth rate by focusing on the dominant term of its running time and ignoring constants and lower-order terms. 
However, in practice, these constants can significantly impact performance on real computers. 
In this section, we present an analysis of the constants in the BMSSP algorithm to allow a more precise comparison with Dijkstra's algorithm. 
(For simplicity, we focus only on the dominant term and neglect the others.)

For sparse graphs, Dijkstra's algorithm has time complexity $O(n \log n)$, which can be approximated as
$c_1 \cdot n \log_2 n$,
where $c_1$ is a constant reflecting the implementation overhead.

Similarly, the BMSSP algorithm has complexity $O(n \log^{2/3} n)$, which we approximate as
$c_2 \cdot n \log_2^{2/3} n$,
where $c_2$ is a constant reflecting the implementation overhead.

Since $O(n \log n)$ grows faster than $O(n \log^{2/3} n)$, there exists a threshold $n_0$ beyond which BMSSP will outperform Dijkstra. 
Formally, we define
\begin{equation}    
n_0 = \min \set{n \mid c_2 \, n \, \log_2^{2/3} n < c_1 \, n \, \log_2 n},
\end{equation}

which simplifies to
\begin{equation}
n_0 = \min \set{n \mid \log_2 n > (c_2 / c_1) ^ 3}.
\end{equation}

Assuming different ratios $c_{2} / c_{1}$ (that is, assuming different rates at which the BMSSP constant is greater than Dijkstra's), the threshold values $n_0$ are as follows:

\begin{table}[!htpb]
\centering
\begin{tabular}{cc}
\hline
 $\mathbf{c_2 / c_1}$ & $\mathbf{n_0}$  \textbf{(approx.)} \\ \hline
2  & $256$                \\ 
3  & $10^{9}$       \\
4  & $10^{20}$      \\ 
5  & $10^{38}$      \\ 
7  & $10^{103}$     \\ 
10 & $10^{301}$     \\ \hline
\end{tabular}
\label{tab:bmssp-performance}
\end{table}

This analysis predicts that even if the constant in a BMSSP implementation is only five times larger than that of Dijkstra's algorithm, Dijkstra will remain faster for almost all practical graph sizes.

\section{Implementation Details}\label{sec:implementation_details}

We provide implementations of the algorithms proposed by \citet{duan2025breaking} and
\citet{dijkstra1959note}. All implementations are written in \texttt{C++20} and compiled
with \texttt{g++} using the \texttt{-O3} optimization flag. 
The code is publicly available
at \texttt{\href{https://github.com/lcs147/bmssp}{https://github.com/lcs147/bmssp}}.

Dijkstra's algorithm is implemented using the binary heap provided by the \texttt{C++}
standard library (\texttt{std::priority\_queue}). We adopt this data structure because
binary heaps are often more efficient in practice than Fibonacci heaps
\citep{cherkassky1996shortest}. As a result, the Dijkstra implementation used in our
experiments has worst-case time complexity $O((n+m)\log n)$, when restricted to sparse
graphs ($m = O(n)$), this simplifies to $O(n \log n)$.

The algorithm proposed by \citet{duan2025breaking} requires that all simple paths have distinct lengths. To satisfy this requirement in a general setting, we must break ties consistently when paths have equal length. As described in the original paper, we represent the estimated distance to a vertex $v$ as the tuple:

\begin{equation}    
\hat{d}[v] = \big( \text{length}[v],\ \text{hops}[v],\ v,\ \text{pred}[v] \big).
\end{equation}

Here, $\text{length}[v]$ is the total weight of the current shortest path to $v$, $\text{hops}[v]$ is the number of edges in this path, and $\text{pred}[v]$ is the predecessor of $v$ in this path. Paths are compared using the lexicographic order on these tuples, which guarantees a unique ordering even if their $\text{length}$ values are identical.

We implement the algorithm of \citet{duan2025breaking} (BMSSP) following the specifications
of its three subroutines and its core data structure. In our experimental study, we
evaluate three implementations of BMSSP that differ in how the theoretical guarantees
are realized.

The \textbf{BMSSP-WC} implementation is designed to preserve the theoretical worst-case
time complexity of $O(m \log^{2/3} n)$. The \textbf{BMSSP-CD} implementation also
guarantees a worst-case time complexity of $O(m \log^{2/3} n)$; however, before
computing shortest paths, it applies a preprocessing step that transforms the input
graph into a constant-degree graph in which each vertex has out-degree at most $2$.
This transformation is required to achieve the stated
worst-case bound \citep{duan2025breaking}.

While the BMSSP-WC implementation is fully deterministic, the \textbf{BMSSP-Expected} implementation relies on randomized data structures and selection algorithms and achieves an expected time complexity of $O(m \log^{2/3} n)$.

In the following, we report implementation details common to all three versions. Then, we report implementation details specific to each version.

\subsection{Common Details}
\paragraph*{\textsc{FindPivots} forest building}

In the \textsc{FindPivots} subroutine, we do not explicitly build the directed forest $F$. Since we only need to identify roots whose trees contain at least $k$ vertices, we maintain an array $\text{root}[v]$ that stores the root of the tree to which vertex $v$ belongs. Whenever an edge $(u,v)$ is relaxed, we update $\text{root}[v] := \text{root}[u]$. This approach allows us to identify the final set of pivots efficiently, without the
overhead of explicitly building the forest or traversing its vertices.

\paragraph*{Set operations}

To guarantee $O(1)$ worst-case set operations 
(including unions, membership tests, and duplicate removal), we consistently use global direct-address tables (DATs) throughout the implementation.

\paragraph*{Disjointness of $U_i$ sets}

Experimentally, we encountered a subtle discrepancy with Remark~3.8 in \citet{duan2025breaking}, which asserts that the sets $U_i$ in a recursive call of \textsc{BMSSP} (Algorithm~3) are disjoint. This property may fail: Consider a vertex $x$ that enters the data structure $D$ with an initial distance $d_{\text{old}}$. If $x$ is later marked as complete by a recursive call with a shorter distance $d_{\text{new}}$ (so that $x \in U_i$) but is never extracted from $D$, it remains in $D$ with the outdated key $d_{\text{old}}$. When $x$ is eventually extracted (still with key $d_{\text{old}}$), it becomes a frontier vertex, causing its shortest-path subtree to be rediscovered with the correct distance $d_{\text{new}}$. Consequently, $x$ and its subtree may be assigned to another set $U_j$, violating disjointness.

We restore the disjointness property by removing a vertex from $D$ as soon as it is marked complete; more specifically, before line 15 of Algorithm 3, we perform $D.\mathtt{erase}(u)$.\footnote{We have consulted with the original authors, who confirmed both the problem and our proposed solution.}

Let $N$ be the total number of elements and $M$ be the maximum number of elements in a block. To implement this, we can check if the element is present in $O(1)$ time using set operations. If it is not present, do nothing; otherwise, erase the element from its block. If the block becomes empty, erase the block itself. Erasing a block takes $O(\log (N / M))$ time using a binary search tree, while all other operations are set-based and take $O(1)$ time. This modification satisfies Remark 3.8 without increasing the asymptotic complexity because the deletion is amortized against the insertion.

\paragraph*{Choice of $l$, $k$ and $t$}
We set $k = \floor{\log_2^{1/3} n}$ and $t = \floor{\log_2^{2/3} n}$. The algorithm is then called with $l = \ceil{\log_2 (n) / t}$, following the parameters in the original paper to achieve the same complexity.

\paragraph*{Correctness}
We validated our implementations against Dijkstra’s algorithm across all datasets. Additionally, the public repository uses automated testing on small-scale instances to verify the integrity of future modifications.

\subsection{BMSSP-WC Specific Details}


\paragraph*{Additional memory usage}
The block-based data structure in the original paper stores at most $N$ key-value pairs. It must support finding any key and its associated data in $O(1)$ time.

Instead of using a hash table to find the keys (which provides expected $O(1)$ operations), we employ global direct-address tables (DATs) of size $O(n)$, where $n$ is the number of vertices. Naively, each recursive call of \textsc{BMSSP} (Algorithm~3) would require its own DAT, leading to an infeasible number of tables. However, due to the sequential nature of the divide-and-conquer recursion, we can reuse DATs across calls. Specifically, we maintain only one DAT \emph{per recursion level}, as recursive calls at the same level are processed sequentially and can share the same memory. With $O(\log^{1/3} n)$ recursion levels and a DAT of size $\Theta(n)$ per level, the total space complexity becomes $\Theta(n \log^{1/3} n)$.

\paragraph*{Faster $O(n)$ Selection}

In the block-based linked-list data structure described by \citet{duan2025breaking}, all
operations (\textbf{Insert}, \textbf{Batch Prepend}, and \textbf{Pull}) rely on selecting
the $k$-th smallest element as a subroutine, which must run in $O(n)$ worst-case time.

Rather than using \textsc{QuickSelect}~\citep{hoare1961algorithm} combined with the
\textsc{Median of Medians} algorithm \citep{blum_time_1973}, as suggested in the original
paper, we adopt \textsc{QuickSelectAdaptive} proposed by \citet{alexandrescu2017fast}.
This algorithm preserves the deterministic $O(n)$ worst-case time complexity while
achieving significantly smaller constant factors in practice. We use the implementation
of \textsc{QuickSelectAdaptive} provided by \citet{miniselect2020}, which is incorporated
into our code. According to the benchmarks reported in \citet{miniselect2020},
\textsc{QuickSelectAdaptive} is approximately $10\times$ faster than
the Median of Medians algorithm.

\vspace{1em}

\subsection{BMSSP-Expected Specific Details}

\paragraph*{No additional memory usage}

Unlike \textbf{BMSSP-WC}, this variant allows operations with expected-time guarantees
instead of worst-case bounds. Consequently, we use hash tables within the block-based
data structure to store keys and their associated data in expected $O(1)$ time.

Since the sets $U_i$ are pairwise disjoint (Remark~3.8), the total number of complete vertices $|\bigcup U_i|$ in the recursion active stack is $O(n)$. Given the constant-degree assumption, each vertex contributes only a constant number of adjacent elements to the priority queue. Therefore, all priority queues collectively contain at most $O(n)$ elements at any point, resulting in a total space complexity of $O(n)$.

We use an efficient hash map based on Robin Hood hashing with backward shift deletion,
available at
\texttt{\href{https://github.com/martinus/unordered_dense}{https://github.com/martinus/unordered\_dense}}.

\paragraph*{Floyd-Rivest Selection}

For selecting the $k$-th smallest element within the block-based data structure, we use
the algorithm of \citet{floyd1975expected}, which has expected linear time complexity and
performs $n + \min(n-k, k) + o(n)$ comparisons. We adopt the implementation provided by
\citet{miniselect2020}, which is integrated into our code. According to the benchmarks
reported in \citet{miniselect2020}, Floyd--Rivest selection is the fastest method in
nearly all tested scenarios.

\section{Experiments}\label{sec:experiments}
 
All experiments were conducted on a computer with \texttt{32~GB} of memory 
and an \texttt{Intel Core i5-10400F} processor running at \texttt{2.90~GHz}, 
under \textit{Linux Mint 21.3 Cinnamon}.

\subsection{Graphs}

We evaluate the algorithms on three classes of instances:

\begin{itemize}
    \item \textbf{Sparse Random Graphs:}
    Randomly generated directed graphs with vertex set sizes of
    $2^7, 2^8, \dots, 2^{25}$.
    Vertices are labeled from $1$ to $n$.
    The code used to generate these instances is available in the repository.

    \begin{itemize}
        \item \textbf{D3}: Graphs with mean out-degree $3$ and maximum out-degree $4$.
        It is guaranteed that vertex $1$ can reach all other vertices.
        
        \item \textbf{H3}: Graphs with mean out-degree $3$ and no restriction on the
        maximum out-degree.
        These graphs are strongly connected; that is, every vertex can reach every other vertex.

    \end{itemize}
    
    \item \textbf{Road Graphs (USA):}
    Twelve graphs representing road networks from different regions of the United States,
    obtained from the 9th DIMACS Implementation Challenge~\citep{dimacs}.
    Edge weights correspond to the average travel time along each road.
    As is typical for road networks, all graphs are sparse, and vertices have small degrees.
    The number of vertices and edges ranges from approximately
    $10^5$ to $10^7$.

    \item \textbf{Grid Graphs:}
    Generated with vertex set sizes of $2^8, \dots, 2^{24}$.
    Vertices correspond to points with integer coordinates $(x,y)$,
    where $1 \leq x \leq R$, $1 \leq y \leq C$, and $|V| = R \times C$.
    Each vertex has up to eight neighbors:
    $(x \pm 1, y)$, $(x, y \pm 1)$,
    $(x \pm 1, y \pm 1)$.
    The code used to generate these instances is available in the repository.

    \begin{itemize}
        \item \textbf{SGridED}: Square grids with $R = C$ and Euclidean edge weights.
        Horizontal and vertical edges have weight $1$, while diagonal edges have weight $\sqrt{2}$.
        
        \item \textbf{RGridED}: Rectangular grids with $R = 4C$ and Euclidean edge weights.
        Horizontal and vertical edges have weight $1$, while diagonal edges have weight $\sqrt{2}$.
        
        \item \textbf{SGridR}: Square grids with $R = C$ and edge weights drawn uniformly at random
        from the interval $[1, 10^5]$.
        
        \item \textbf{RGridR}: Rectangular grids with $R = 4C$ and edge weights drawn uniformly at random
        from the interval $[1, 10^5]$.
    \end{itemize}
\end{itemize}

All randomly generated instances used fixed random seeds, enabling reproducibility.

\subsection{Experimental Evaluations}
We conducted three experimental evaluations. In Experiment~1, we compare the performance of the BMSSP-WC implementation with Dijkstra's algorithm~\citep{dijkstra1959note}.
Both algorithms are executed on instances D3, H3, USA, and Grid Graphs.
Since these graphs are sparse and do not exhibit high maximum out-degrees
(with the exception of the H3 instance), degree normalization is not necessary.
In Experiment~2, we evaluate the impact of degree normalization on performance by running BMSSP-WC and BMSSP-CD on instances without restrictions on the maximum out-degree, namely USA and H3. In Experiment~3, we compare the worst-case $O(m \log^{2/3} n)$ implementation with the expected-time $O(m \log^{2/3} n)$ implementation.
BMSSP-WC and BMSSP-Expected are executed on instances D3, H3, USA, and Grid Graphs.

For all instances, shortest paths are computed using vertex $1$ as the source. We report the average wall-clock time across five independent runs per instance to mitigate the impact of system noise and timing variability. This average, measured in milliseconds, serves as our primary performance metric. The results are detailed in \Cref{sec:results1,sec:results2,sec:results3}.

\section{Experiment 1 - BMSSP-WC vs. Dijkstra}\label{sec:results1}

In Experiment~1, we compare the performance of BMSSP-WC and Dijkstra's algorithm
on three classes of instances: Random Graphs, Road Graphs, and Grid Graphs.

\subsection{Random Graphs}

\begin{figure}[!htpb]
    \begin{adjustwidth}{-1.5cm}{-1.5cm}
        \centering
        \begin{subfigure}{0.31\linewidth}
            \centering
            \includegraphics[width=\linewidth] {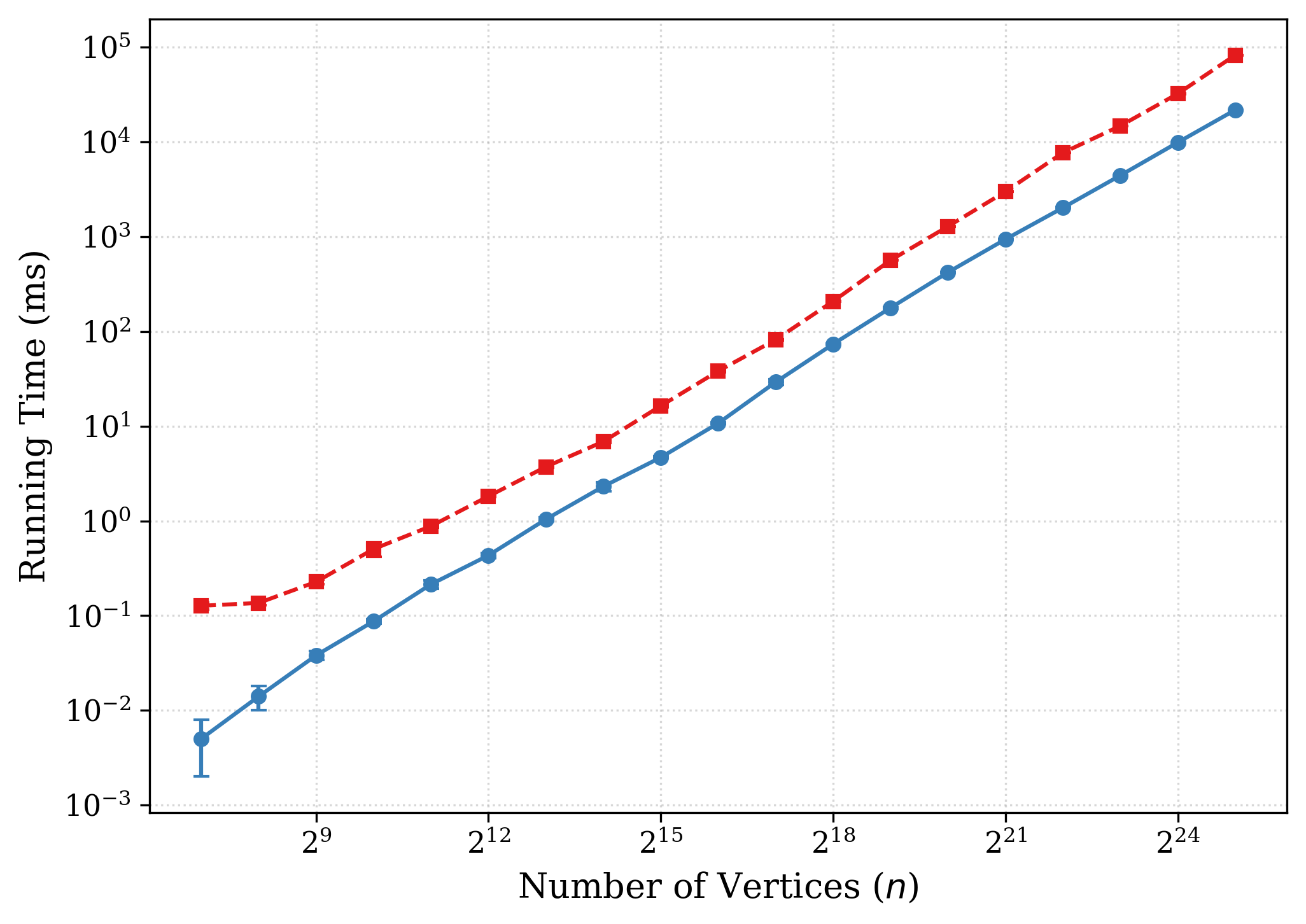}
            \caption{Random sparse graphs D3.}
            \label{fig:times_wc_vs_dij_d3}
        \end{subfigure}
        \hfill
        \begin{subfigure}{0.31\linewidth}
            \centering
            \includegraphics[width=\linewidth]{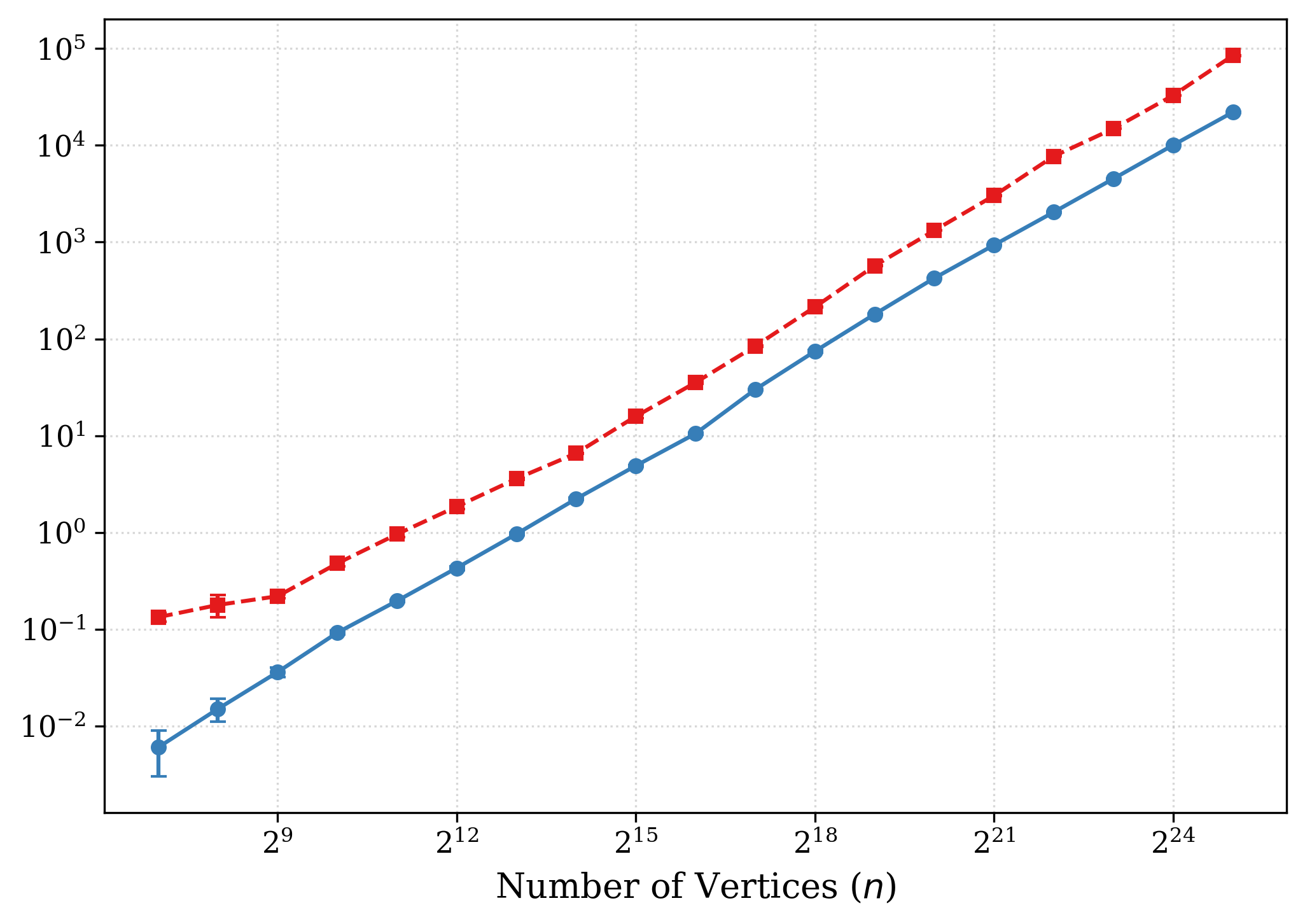}
            \caption{Random sparse graphs H3.}
            \label{fig:times_wc_vs_dij_h3}
        \end{subfigure}
        \hfill
        \begin{subfigure}{0.31\linewidth}
            \centering
            \includegraphics[width=\linewidth]{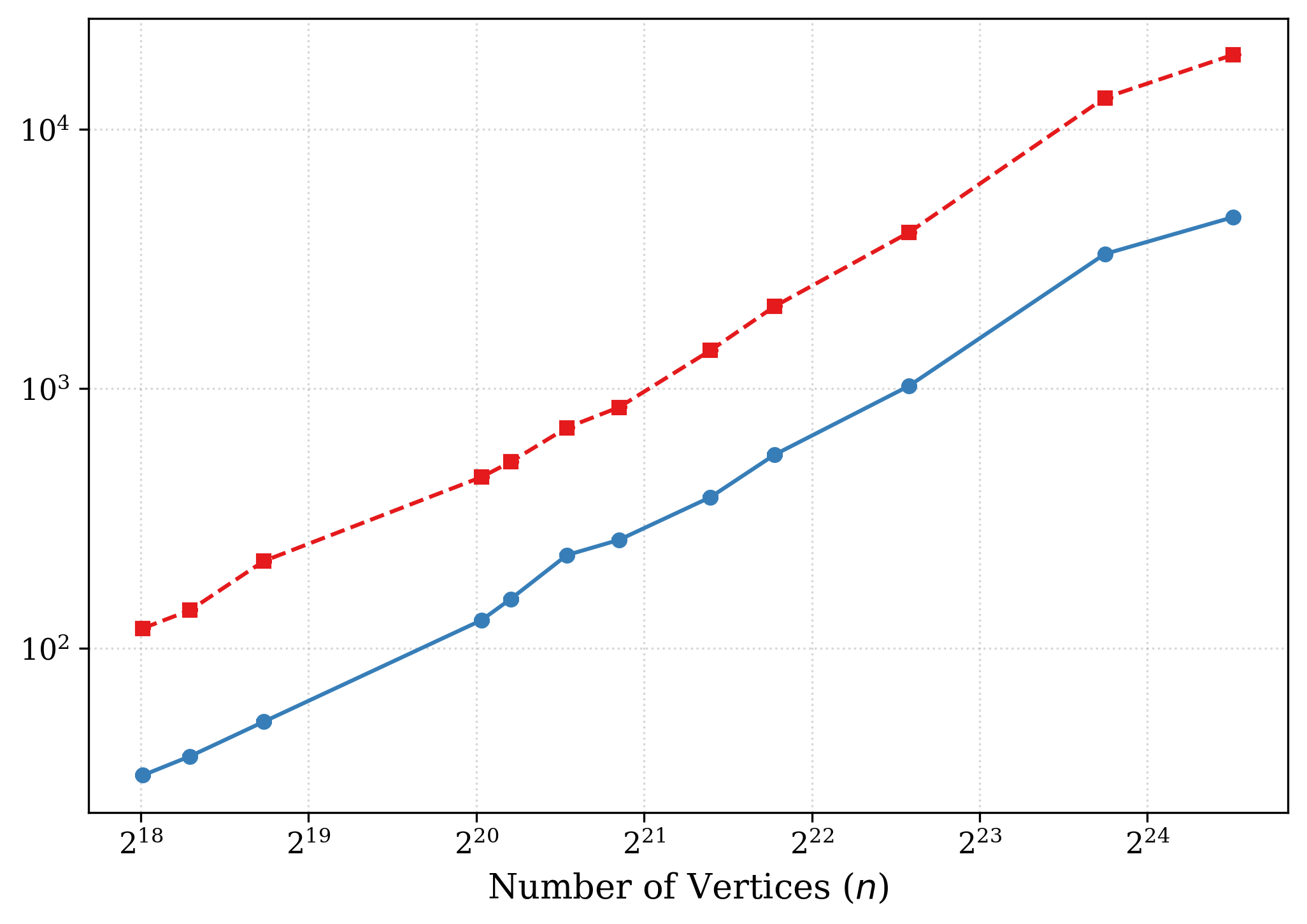}
            \caption{USA road networks.}
            \label{fig:times_wc_vs_dij_usa}
        \end{subfigure}
        
        \vspace{0.1cm}

        \begin{subfigure}{0.31\linewidth}
            \centering
            \includegraphics[width=\linewidth]{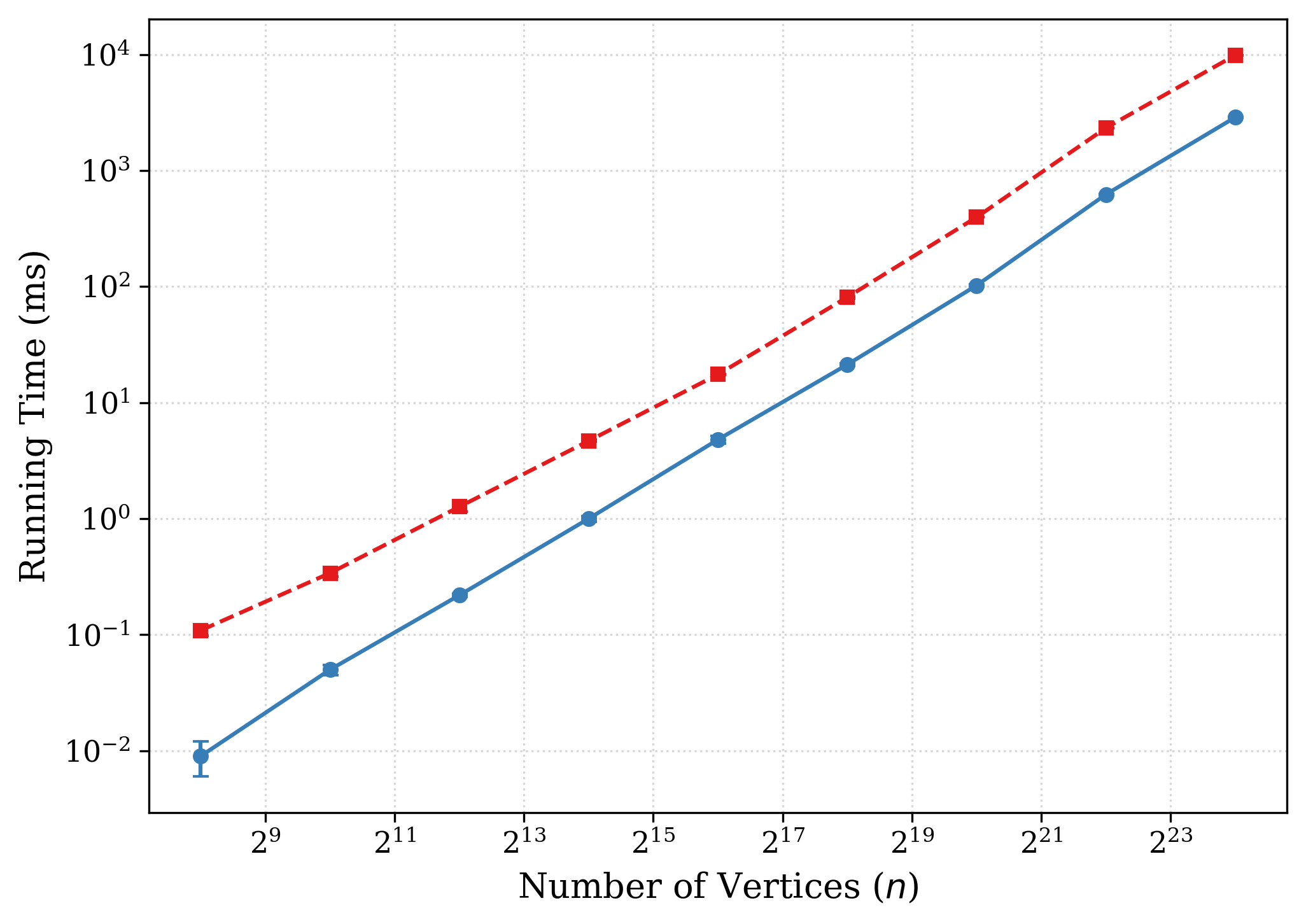}
            \caption{Square grid graphs (SGridED).}
            \label{fig:times_wc_vs_dij_sgrided}
        \end{subfigure}
        \hspace{1cm} 
        \begin{subfigure}{0.31\linewidth}
            \centering
            \includegraphics[width=\linewidth]{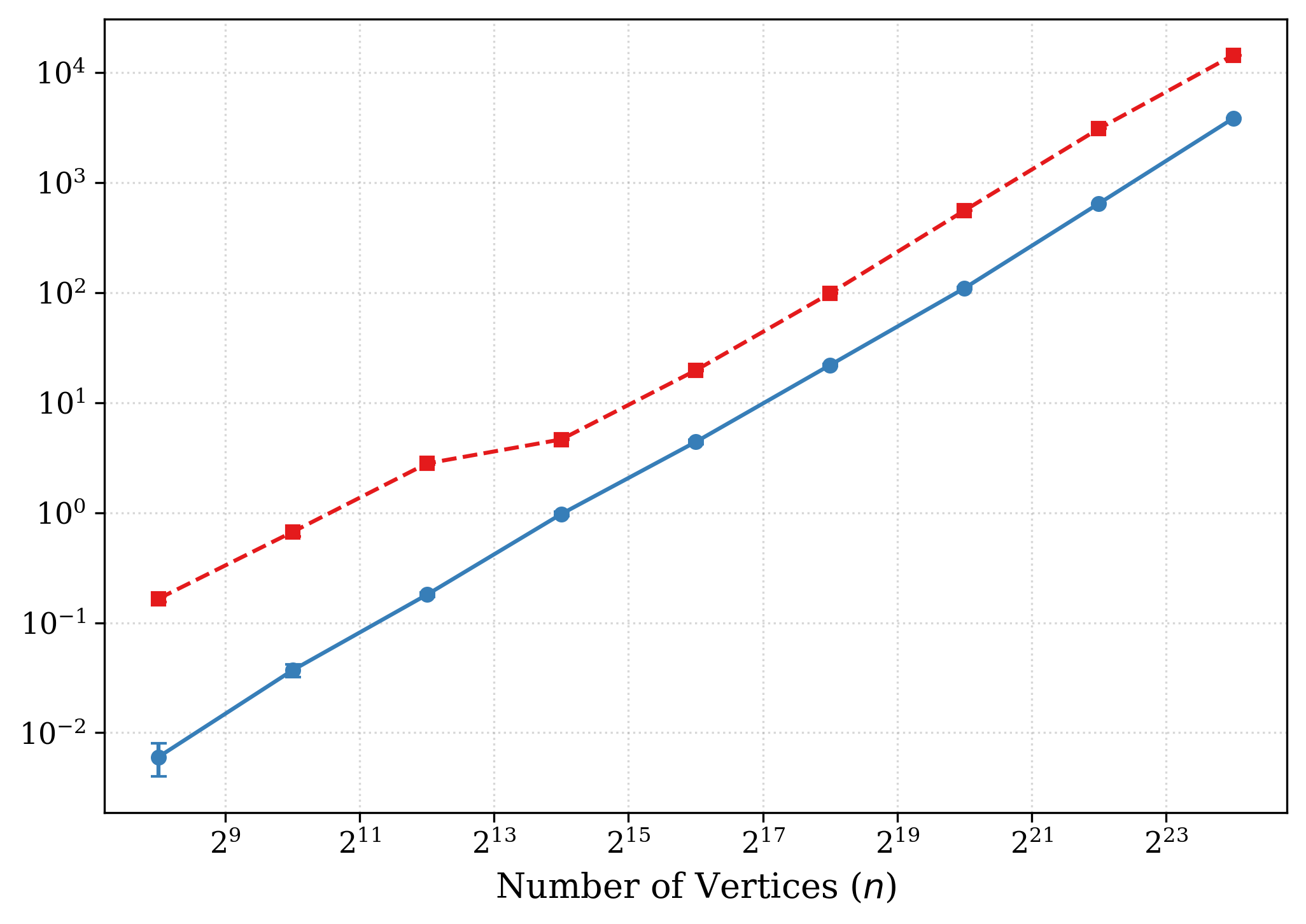}
            \caption{Rectangular grid graphs (RGridED).}
            \label{fig:times_wc_vs_dij_rgrided}
        \end{subfigure}

        \vspace{0.1cm}

        \begin{subfigure}{0.31\linewidth}
            \centering
            \includegraphics[width=\linewidth]{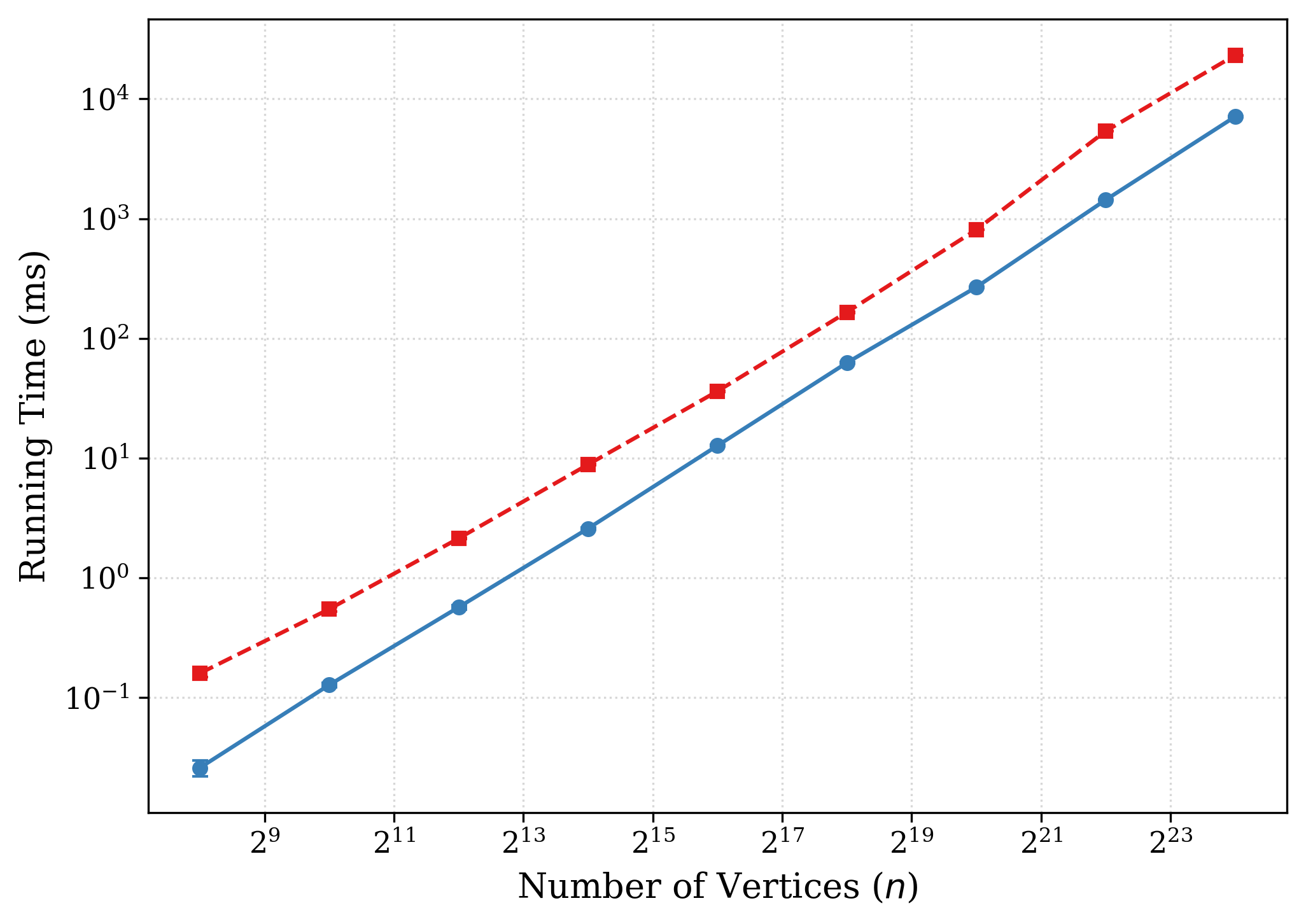}
            \caption{Square grid graphs (SGridR).}
            \label{fig:times_wc_vs_dij_sgridr}
        \end{subfigure}
        \hspace{1cm}
        \begin{subfigure}{0.31\linewidth}
            \centering
            \includegraphics[width=\linewidth]{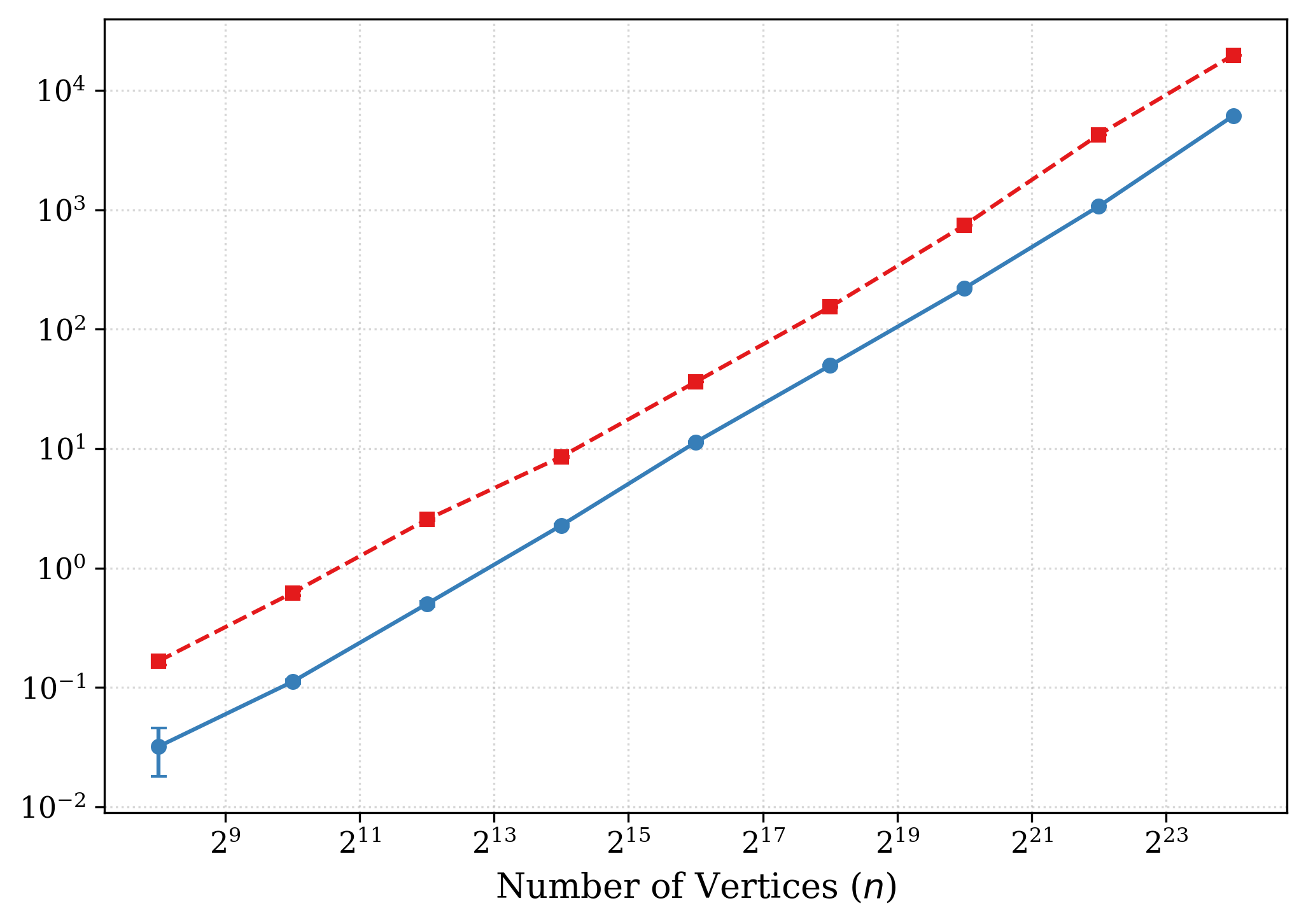}
            \caption{Rectangular grid graphs (RGridR).}
            \label{fig:times_wc_vs_dij_rgridr}
        \end{subfigure}

        \vspace{0.1cm}

        \centering
        \begin{tikzpicture}
            \draw[colorDijkstra, line width=1.5pt] (0,0) -- (0.5,0) node[right, black] {\small Dijkstra};
            \begin{scope}[shift={(2.5,0)}]
                \draw[colorBMSSP, line width=1.5pt] (0,0) -- (0.5,0) node[right, black] {\small BMSSP};
            \end{scope}
        \end{tikzpicture}

        \caption{Average execution time of BMSSP-WC and Dijkstra as a function of the number of vertices on a log-log scale.}
        \label{fig:times_wc_vs_dijkstra}
    \end{adjustwidth}
\end{figure}

\Cref{fig:times_wc_vs_dij_d3,fig:times_wc_vs_dij_h3} report the running times of both algorithms on
random graph instances of types D3 and H3.
The observed behavior is similar for both instance types.
Across all tested instances, Dijkstra's algorithm consistently outperforms BMSSP-WC.
However, the performance gap between the two algorithms decreases as the graph size increases.

From a theoretical perspective, one would expect the running-time curves of the two
algorithms to eventually intersect, after which BMSSP-WC becomes faster.
Nonetheless, for graphs with up to $2^{25}$ vertices, this crossover point is not observed.

As shown in \Cref{tab:ex1_performance}, the best observed case for BMSSP-WC
occurs on instances \texttt{random131072D3} and \texttt{H3random131072},
each with $2^{17}$ vertices, where BMSSP-WC is approximately $2.8\times$ slower
than Dijkstra's algorithm.
On average, considering instances with more than $10^3$ vertices,
BMSSP-WC is $3.56\times$ slower than Dijkstra.
For the largest random graphs with $2^{25}$ vertices,
BMSSP-WC computes the shortest paths in approximately $84$ seconds,
compared to $22$ seconds for Dijkstra's algorithm.

\paragraph*{Implementation Check}
To verify whether the implementation follows the expected asymptotic behavior,
we compare the empirical performance ratio with the ratio predicted by the
theoretical time complexities.
Based on the asymptotic bounds of the algorithms, the expected ratio between the
running times of BMSSP and Dijkstra is

\begin{equation}
    f(n) = \frac{\text{Time(BMSSP)}}{\text{Time(Dijkstra)}} = \frac{c_2 \cdot n \log^{2/3} n}{c_1 \cdot n \log n} = \frac{c_2}{c_1} \log^{-1/3} n.
\end{equation}
    
\begin{figure}[!htpb]
    \begin{adjustwidth}{-1.5cm}{-1.5cm}
        \centering
        \begin{subfigure}{0.31\linewidth}
            \centering
            \includegraphics[width=\linewidth]{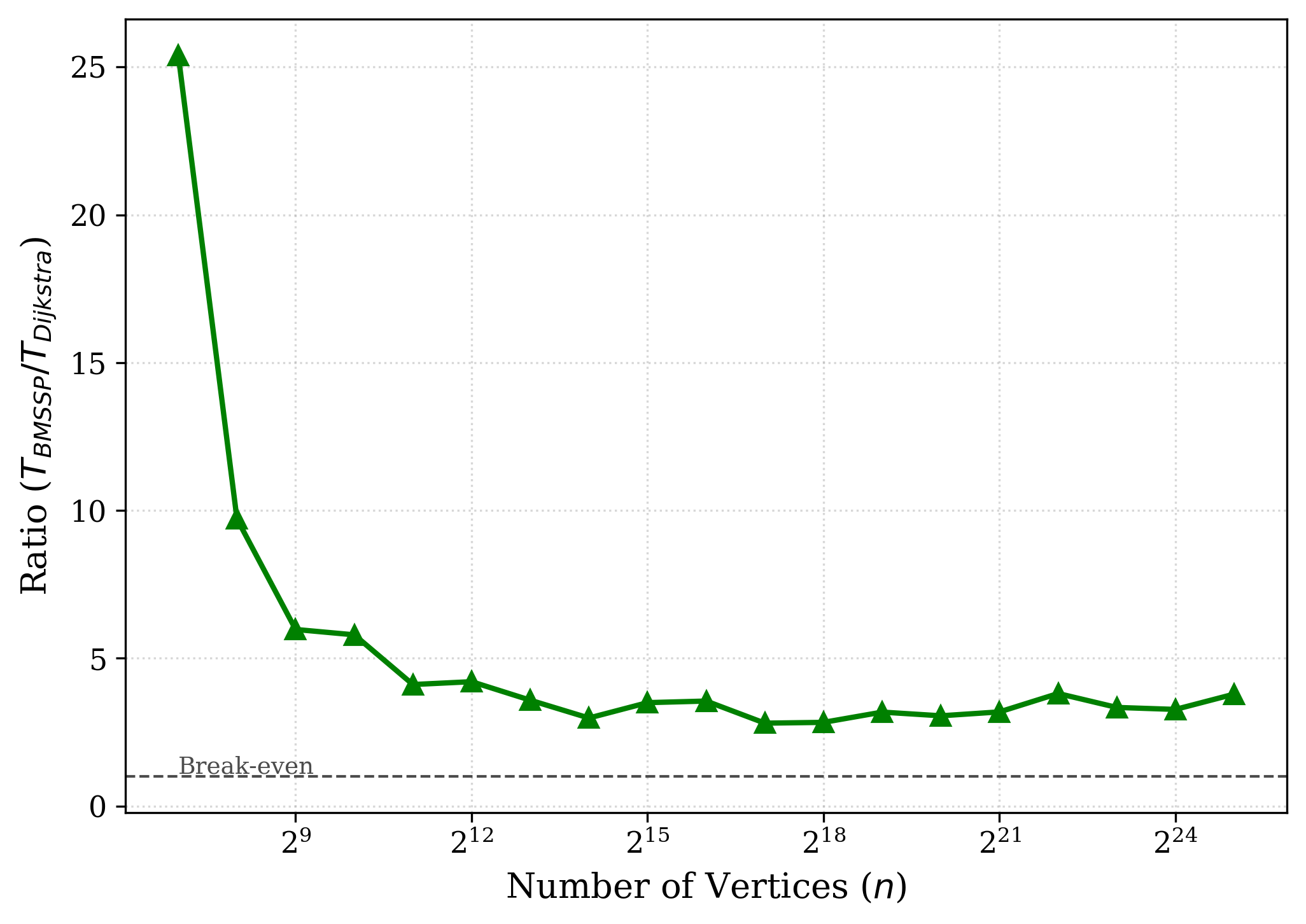}
            \caption{Random sparse graphs D3.}
            \label{fig:ratio_wc_vs_dij_d3}
        \end{subfigure}
        \hfill
        \begin{subfigure}{0.31\linewidth}
            \centering
            \includegraphics[width=\linewidth]{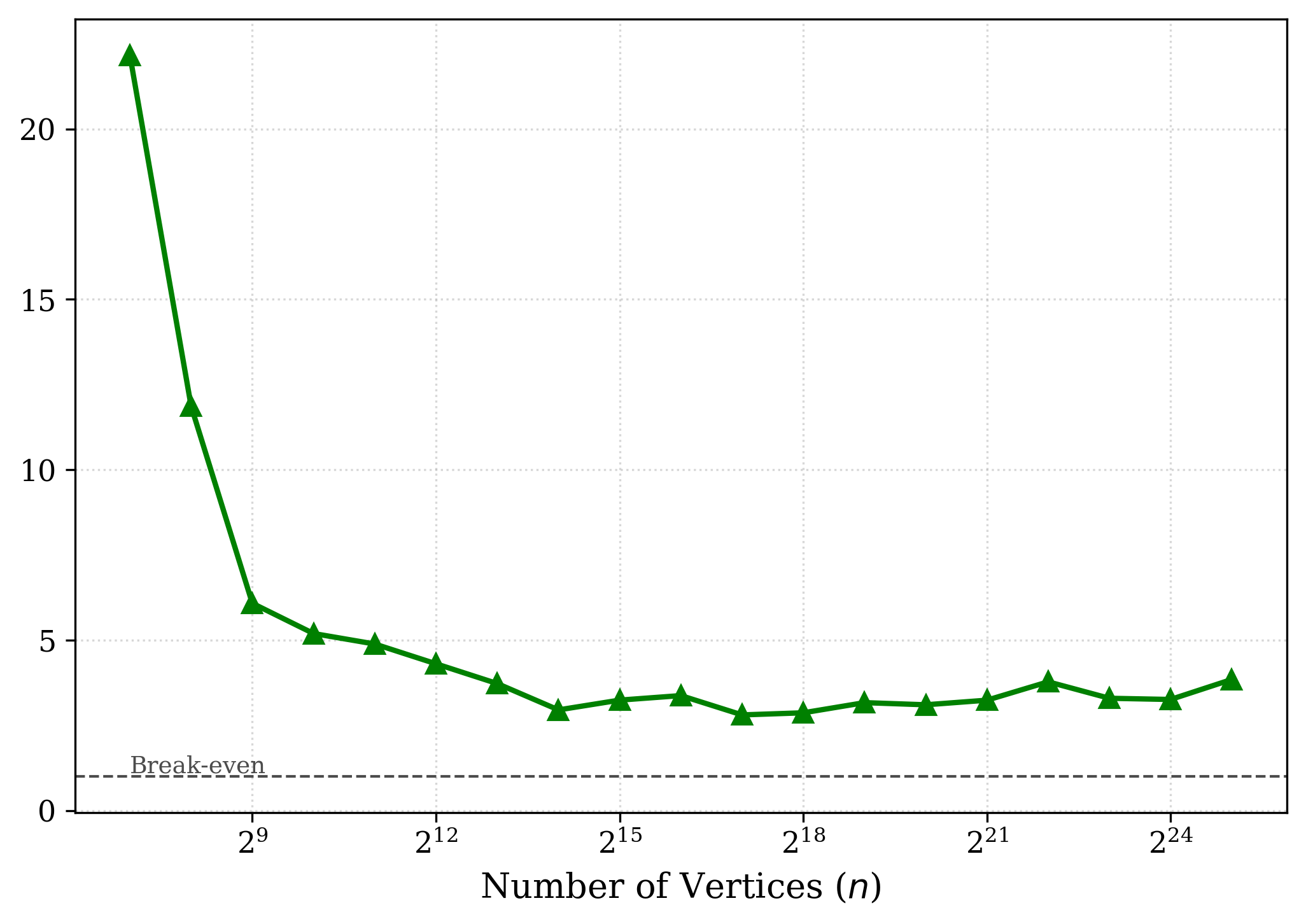}
            \caption{Random sparse graphs H3.}
            \label{fig:ratio_wc_vs_dij_h3}
        \end{subfigure}
        \hfill
        \begin{subfigure}{0.31\linewidth}
            \centering
            \includegraphics[width=\linewidth]{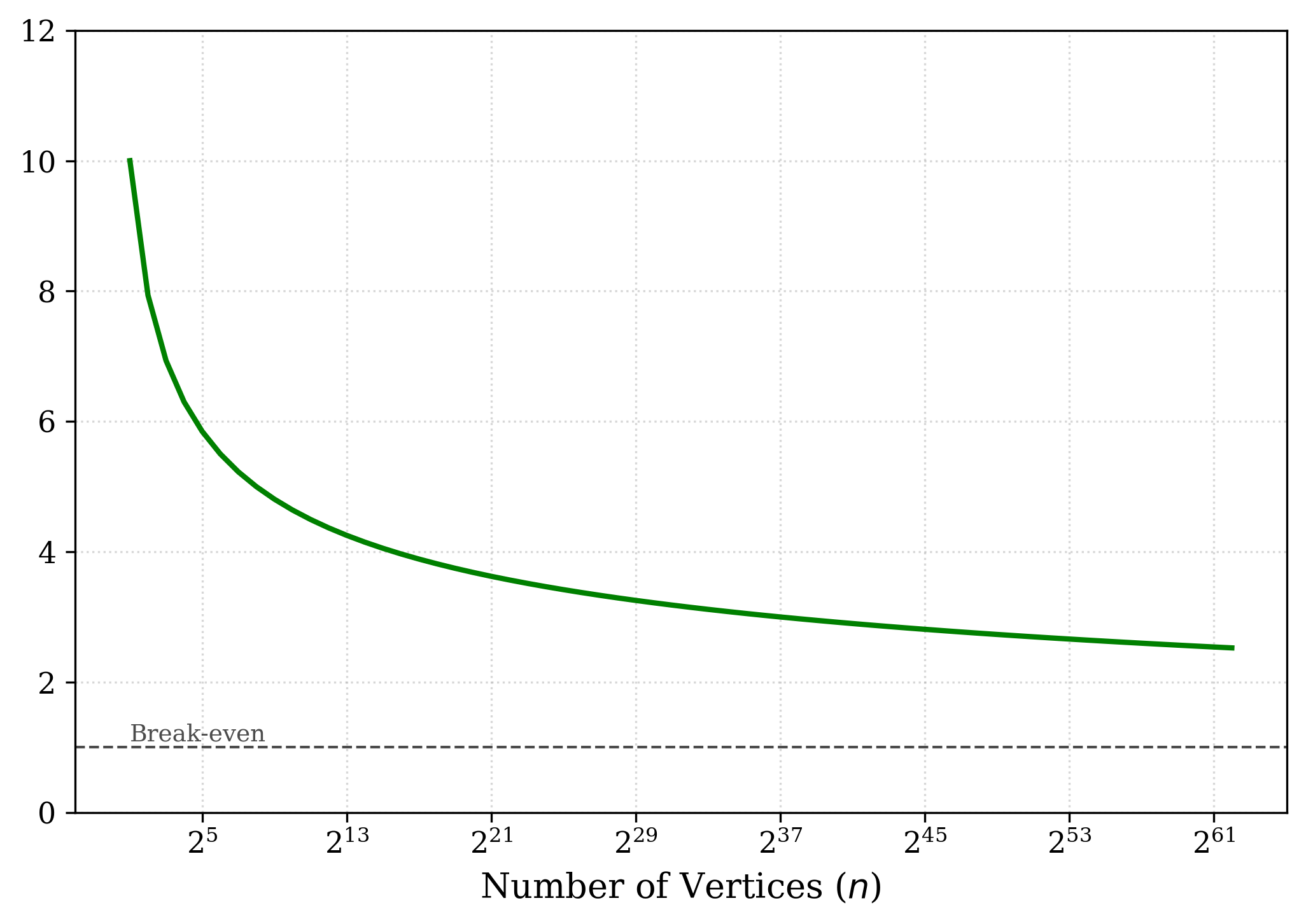}
            \caption{Theoretical ratio with $c_2/c_1 = 10$.}
            \label{fig:theoretical_plot}
        \end{subfigure}
        
        \vspace{0.1cm}

        \begin{subfigure}{0.31\linewidth}
            \centering
            \includegraphics[width=\linewidth]{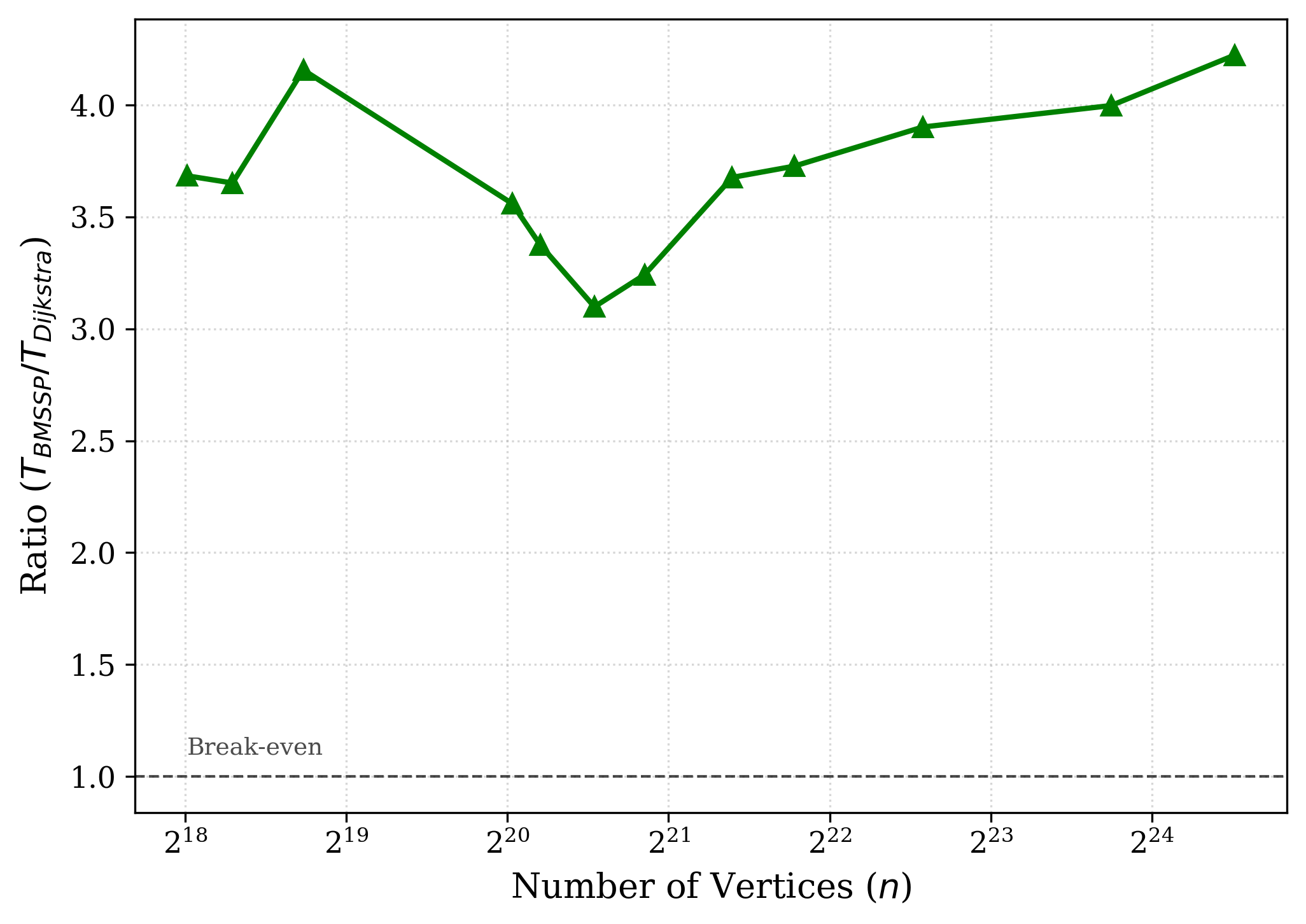}
            \caption{USA road networks.}
            \label{fig:ratio_wc_vs_dij_usa}
        \end{subfigure}
        \hfill
        \begin{subfigure}{0.31\linewidth}
            \centering
            \includegraphics[width=\linewidth]{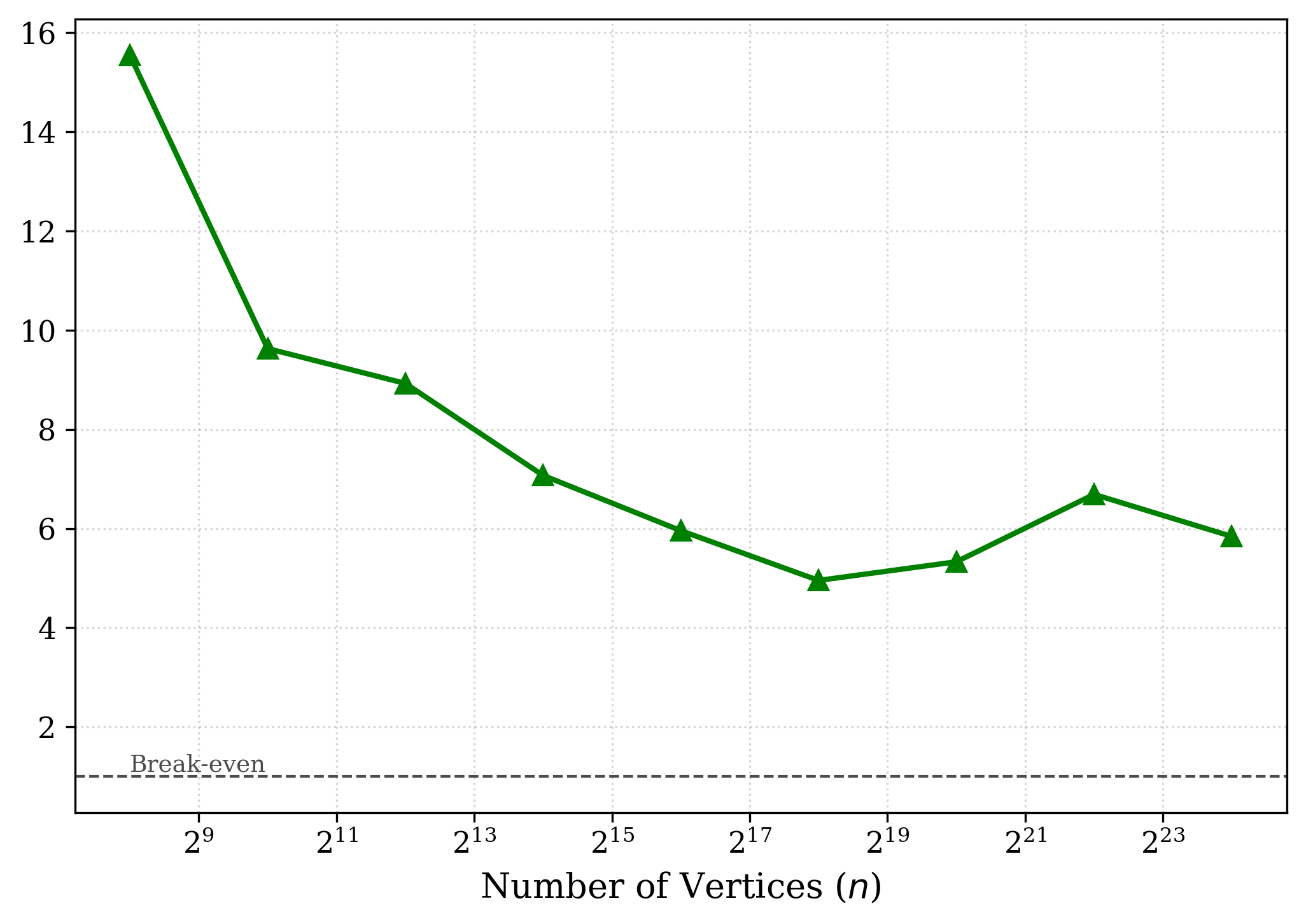}
            \caption{Square grid graphs (SGridED).}
            \label{fig:ratio_wc_vs_dij_sgrided}
        \end{subfigure}
        \hfill
        \begin{subfigure}{0.31\linewidth}
            \centering
            \includegraphics[width=\linewidth]{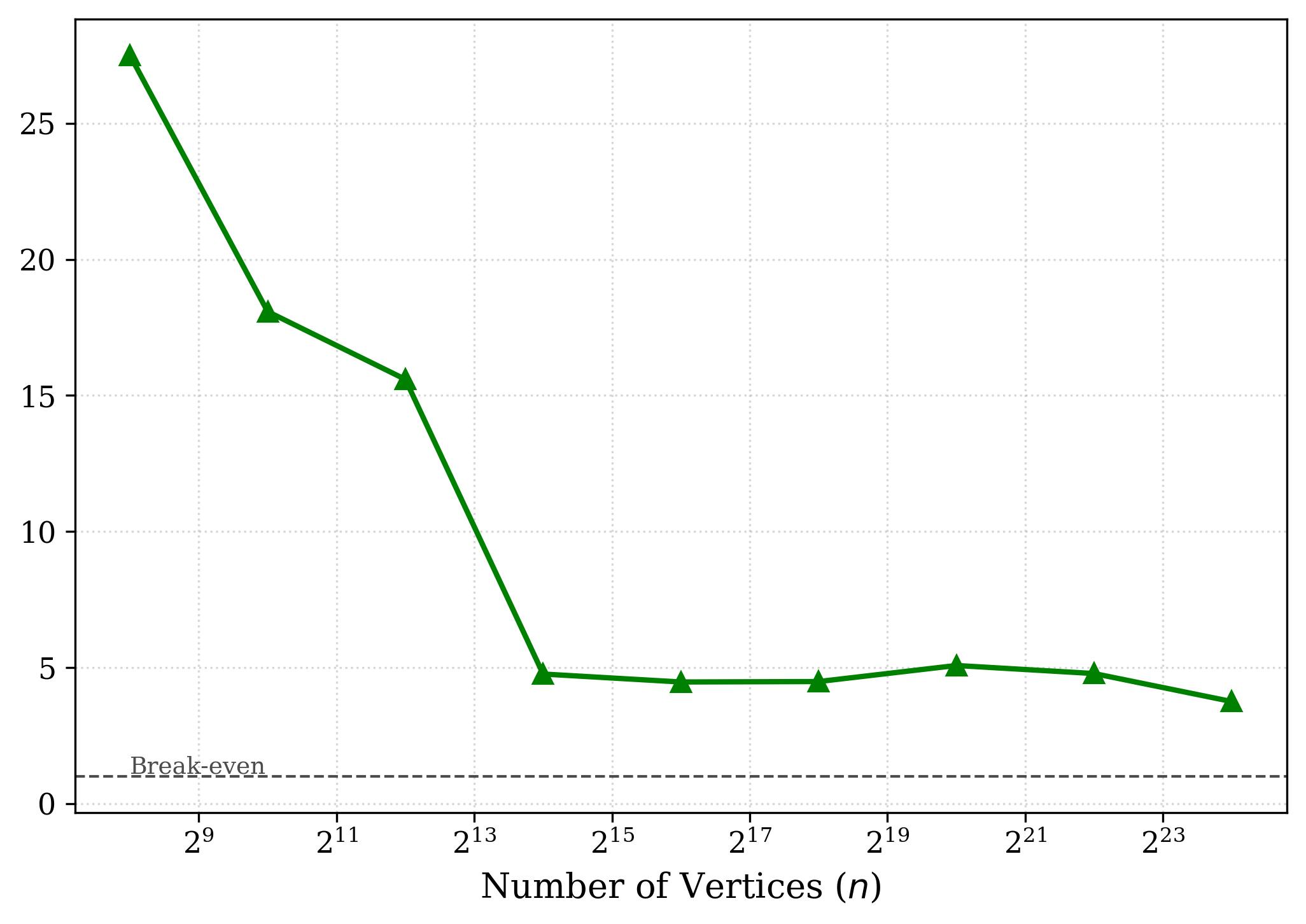}
            \caption{Rectangular grid graphs (RGridED).}
            \label{fig:ratio_wc_vs_dij_rgrided}
        \end{subfigure}

        \vspace{0.1cm}

        \begin{subfigure}{0.31\linewidth}
            \centering
            \includegraphics[width=\linewidth]{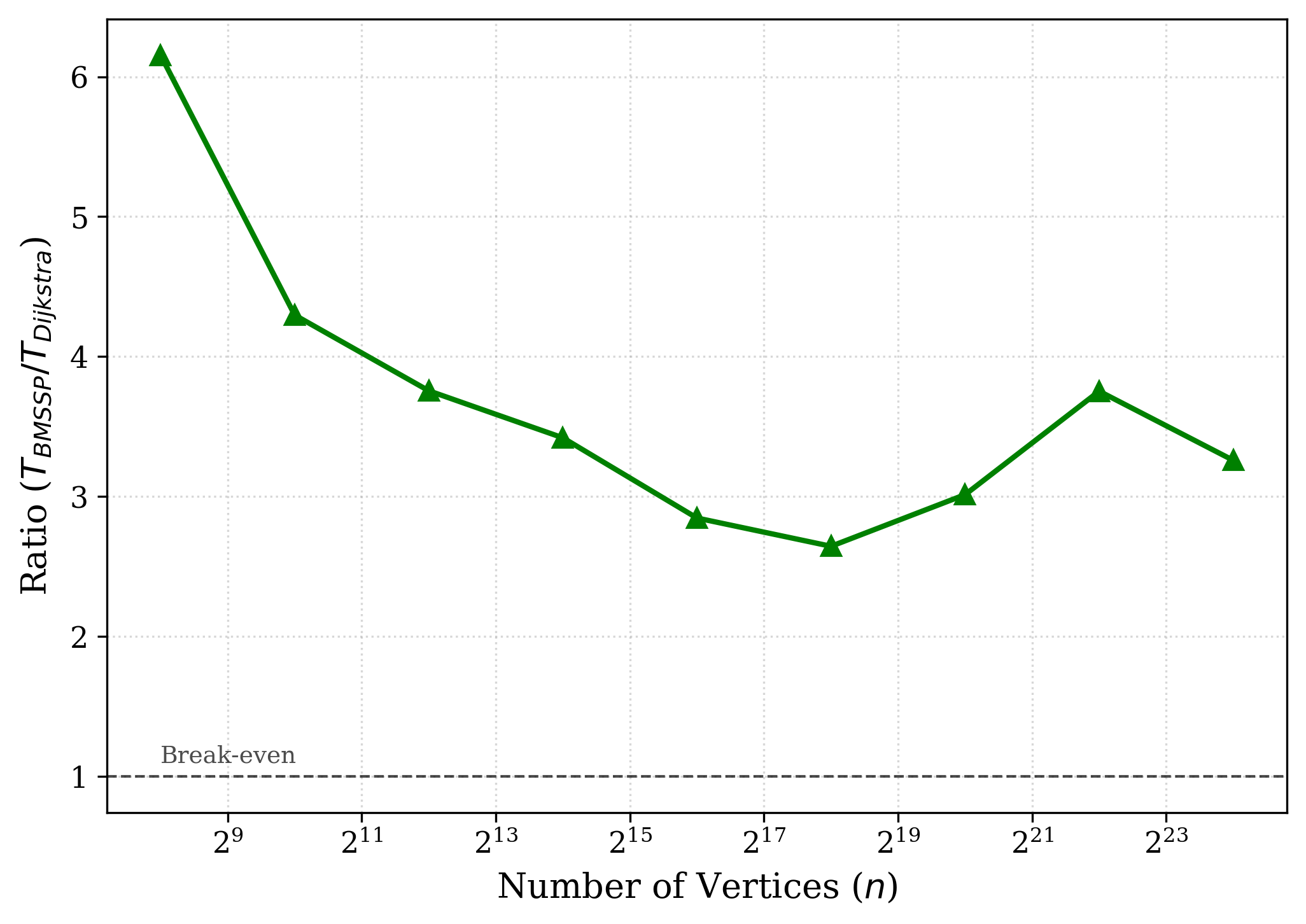}
            \caption{Square grid graphs (SGridR).}
            \label{fig:ratio_wc_vs_dij_sgridr}
        \end{subfigure}
        \hspace{1cm}
        \begin{subfigure}{0.31\linewidth}
            \centering
            \includegraphics[width=\linewidth]{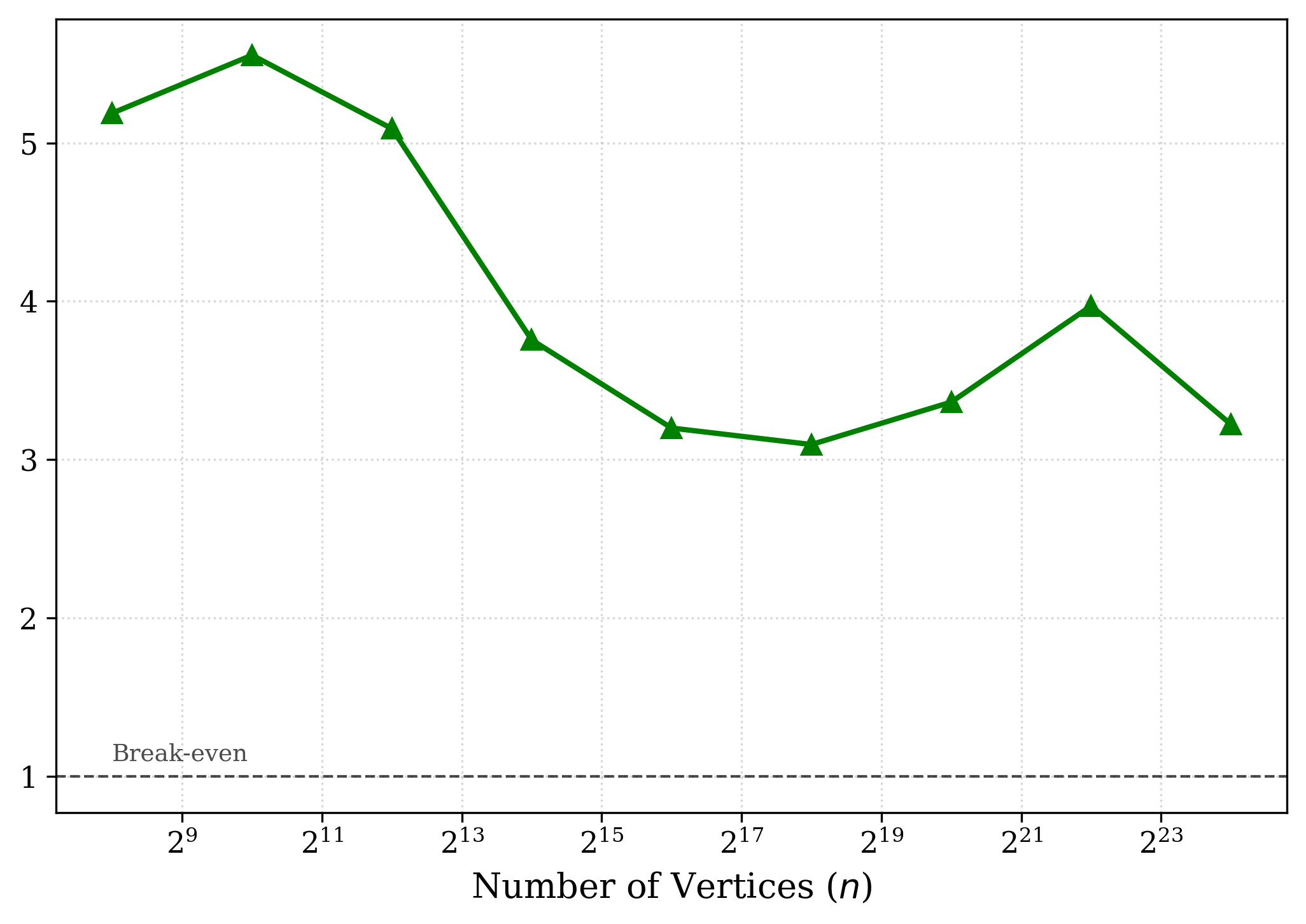}
            \caption{Rectangular grid graphs (RGridR).}
            \label{fig:ratio_wc_vs_dij_rgridr}
        \end{subfigure}

        \caption{Observed ratio between the average running times of 
    BMSSP-WC and
Dijkstra's algorithm.}
        \label{fig:ratio_wc_vs_dij}
    \end{adjustwidth}
\end{figure}

\Cref{fig:theoretical_plot} illustrates the function $f(n)$, which represents the
expected ratio derived from asymptotic analysis.
\Cref{fig:ratio_wc_vs_dij_d3,fig:ratio_wc_vs_dij_h3} report the observed ratios between the average running
times of BMSSP-WC and Dijkstra on random graph instances of types D3 and H3.
The experimental curves closely follow the behavior predicted by $f(n)$, indicating
that the implementation exhibits behavior consistent with the theoretical analysis.

\subsection{Road Graphs (USA)}

\Cref{fig:times_wc_vs_dij_usa} reports the average running times of BMSSP-WC
and Dijkstra's algorithm on the twelve USA road network instances, together with the
observed ratio between their running times as the graph size increases.
Across all instances, Dijkstra consistently outperforms BMSSP-WC.
The performance gap remains relatively stable as the graph size grows, with the
ratio increasing slowly for larger instances.

\Cref{tab:ex1_performance_usa} provides the detailed numerical results.
The observed performance ratios are consistent with those obtained for random
graph instances.
On average, BMSSP-WC is $3.69\times$ slower than Dijkstra.
For the largest instance, representing the full USA road network with approximately
$2.3 \times 10^{7}$ vertices and $5.8 \times 10^{7}$ edges,
BMSSP-WC computes shortest paths in approximately $19$ seconds,
compared to approximately $4$ seconds for Dijkstra's algorithm.

\subsection{Grid Graphs}

\Cref{fig:times_wc_vs_dij_sgridr,fig:times_wc_vs_dij_rgridr,fig:times_wc_vs_dij_rgrided,fig:times_wc_vs_dij_sgrided} report
the average running times of Dijkstra's algorithm and BMSSP-WC on grid graph
instances with Euclidean distances (GridED) and random edge weights (GridR).
In both settings, the results exhibit consistent qualitative behavior across all
tested instances.

For both GridED and GridR graphs, Dijkstra's algorithm consistently outperforms
BMSSP-WC.
As observed in the random graph experiments, the performance gap between the two algorithms decreases as the graph size increases, although no crossover point is
observed within the tested range.

\Cref{tab:ex1_performance_GridED,tab:ex1_performance_GridR}
detail the average running times for GridED and GridR instances, respectively.
For GridED graphs, the smallest instances exhibit a significant performance gap,
with BMSSP-WC being up to $27.5\times$ slower than Dijkstra.
As the grid size increases, this ratio steadily decreases.
For the largest GridED instances with approximately $1.6 \times 10^7$ vertices,
BMSSP-WC is between $3.4\times$ and $3.8\times$ slower than Dijkstra, with running
times reaching approximately $9.9$ seconds for BMSSP-WC and $2.9$ seconds for Dijkstra.

A similar behavior is observed for GridR graphs.
On small instances, BMSSP-WC is between $5\times$ and $6\times$ slower than
Dijkstra, while on larger grids this ratio stabilizes between $3.2\times$ and
$4.0\times$.
For the largest GridR instances, with around $1.6 \times 10^7$ vertices,
BMSSP-WC computes shortest paths in approximately $23$ seconds, compared to
about $7$ seconds for Dijkstra's algorithm.

\Cref{fig:ratio_wc_vs_dij_rgridr,fig:ratio_wc_vs_dij_sgridr,fig:ratio_wc_vs_dij_rgrided,fig:ratio_wc_vs_dij_sgrided} show the observed
ratios between the running times of BMSSP-WC and Dijkstra for GridED and GridR
graphs.
In both cases, the empirical ratios follow a slowly decreasing trend as the number
of vertices increases.

\subsection{Estimating the Threshold Point}\label{ssec:threshold_estimation}

\Cref{tab:threshold_estimation} presents the results of a bootstrap analysis for the estimated threshold point $n_0$ across each graph type. To ensure statistical robustness, we performed 1,000 iterations of resampling with replacement \citep{efron_bootstrap_1979}. In each iteration, we determined the coefficients $c_1$ and $c_2$ by performing a weighted non-linear least squares fit of the models $T_1(n) = c_1 \cdot n \log n$ and $T_2(n) = c_2 \cdot n \log^{2/3} n$ to the resampled execution times \citep{scipy_10_contributors_author_2020}. The optimization weights each data point by the inverse of its measured standard deviation, ensuring more stable measurements exert a greater influence on the final estimates.

The reported values include the mean estimates for $c_1$, $c_2$, and the crossover point $n_0$, alongside their respective 95\% confidence intervals (CI). The results show significant variation according to graph architecture. The minimum mean threshold point is observed for D3 graphs at $10^{212}$, while the maximum mean is $10^{2014}$ for RGridED graphs. The wide confidence intervals of $n_0$ reflect the extreme sensitivity of the $n_0$ calculation to small variations in the fitted constants.

Despite this variance, the practical implications are definitive. Even when considering the most optimistic lower bounds of the 95\% confidence intervals (such as \bestthresholdpoint{} for RGridR), the threshold point $n_0$ remains many orders of magnitude beyond the number of vertices in any feasible graph. This confirms that, while BMSSP is theoretically superior in asymptotic complexity, Dijkstra’s algorithm is significantly more efficient for all practical real-world applications.

\begin{table}[!htpb]
    \centering
    \caption{Threshold point estimation of Dijkstra and BMSSP: Mean [95\% CI]}
    \label{tab:threshold_estimation}
    \small
    \renewcommand{\arraystretch}{2.0}
    \begin{tabular}{lccc}
        \toprule
        \textbf{Graph} & $\mathbf{c_1}$ & $\mathbf{c_2}$ & $\mathbf{n_0}$ \textbf{(approx.)} \\
        \midrule
        H3 & \makecell{$2.06 \times 10^{-5}$ \\ \footnotesize [$1.49 \times 10^{-5}$, $2.32 \times 10^{-5}$]} & \makecell{$1.93 \times 10^{-4}$ \\ \footnotesize [$1.54 \times 10^{-4}$, $2.23 \times 10^{-4}$]} & \makecell{$10^{271}$ \\ \footnotesize [$10^{129}, 10^{579}$]} \\
        \hline
        D3 & \makecell{$2.38 \times 10^{-5}$ \\ \footnotesize [$1.96 \times 10^{-5}$, $2.57 \times 10^{-5}$]} & \makecell{$2.11 \times 10^{-4}$ \\ \footnotesize [$1.52 \times 10^{-4}$, $2.66 \times 10^{-4}$]} & \makecell{$10^{212}$ \\ \footnotesize [$10^{118}, 10^{340}$]} \\
        \hline
        USA & \makecell{$7.59 \times 10^{-6}$ \\ \footnotesize [$6.90 \times 10^{-6}$, $8.33 \times 10^{-6}$]} & \makecell{$7.51 \times 10^{-5}$ \\ \footnotesize [$6.35 \times 10^{-5}$, $8.69 \times 10^{-5}$]} & \makecell{$10^{297}$ \\ \footnotesize [$10^{184}, 10^{429}$]} \\
        \hline
        SGridR & \makecell{$1.49 \times 10^{-5}$ \\ \footnotesize [$1.22 \times 10^{-5}$, $1.67 \times 10^{-5}$]} & \makecell{$1.39 \times 10^{-4}$ \\ \footnotesize [$9.57 \times 10^{-5}$, $1.64 \times 10^{-4}$]} & \makecell{$10^{248}$ \\ \footnotesize [$10^{127}, 10^{374}$]} \\
        \hline
        RGridR & \makecell{$8.36 \times 10^{-6}$ \\ \footnotesize [$4.28 \times 10^{-6}$, $9.54 \times 10^{-6}$]} & \makecell{$7.26 \times 10^{-5}$ \\ \footnotesize [$5.16 \times 10^{-5}$, $9.72 \times 10^{-5}$]} & \makecell{$10^{255}$ \\ \footnotesize [$10^{67}, 10^{550}$]} \\
        \hline
        SGridED & \makecell{$6.55 \times 10^{-6}$ \\ \footnotesize [$5.75 \times 10^{-6}$, $6.80 \times 10^{-6}$]} & \makecell{$1.14 \times 10^{-4}$ \\ \footnotesize [$8.74 \times 10^{-5}$, $1.22 \times 10^{-4}$]} & \makecell{$10^{1607}$ \\ \footnotesize [$10^{931}, 10^{2007}$]} \\
        \hline
        RGridED & \makecell{$5.81 \times 10^{-6}$ \\ \footnotesize [$4.78 \times 10^{-6}$, $6.15 \times 10^{-6}$]} & \makecell{$1.08 \times 10^{-4}$ \\ \footnotesize [$7.84 \times 10^{-5}$, $1.17 \times 10^{-4}$]} & \makecell{$10^{2014}$ \\ \footnotesize [$10^{1095}, 10^{3332}$]} \\
        \hline
        \bottomrule
    \end{tabular}
\end{table}

\section{Experiment 2 - BMSSP-WC vs. BMSSP-CD}\label{sec:results2}

In Experiment~2, we investigate the practical impact of the degree normalization proposed
by \citet{duan2025breaking}. We compare BMSSP-WC against BMSSP-CD on graph
classes that do not have constant degree, namely Road Graphs  and H3 Random Graphs.

\subsection{Road Graphs (USA)}

We first consider the USA road network instances. As real-world street networks, these graphs exhibit very small average
degrees and a low maximum degree.

\begin{figure}[!htpb]
    \includegraphics[width=\linewidth]{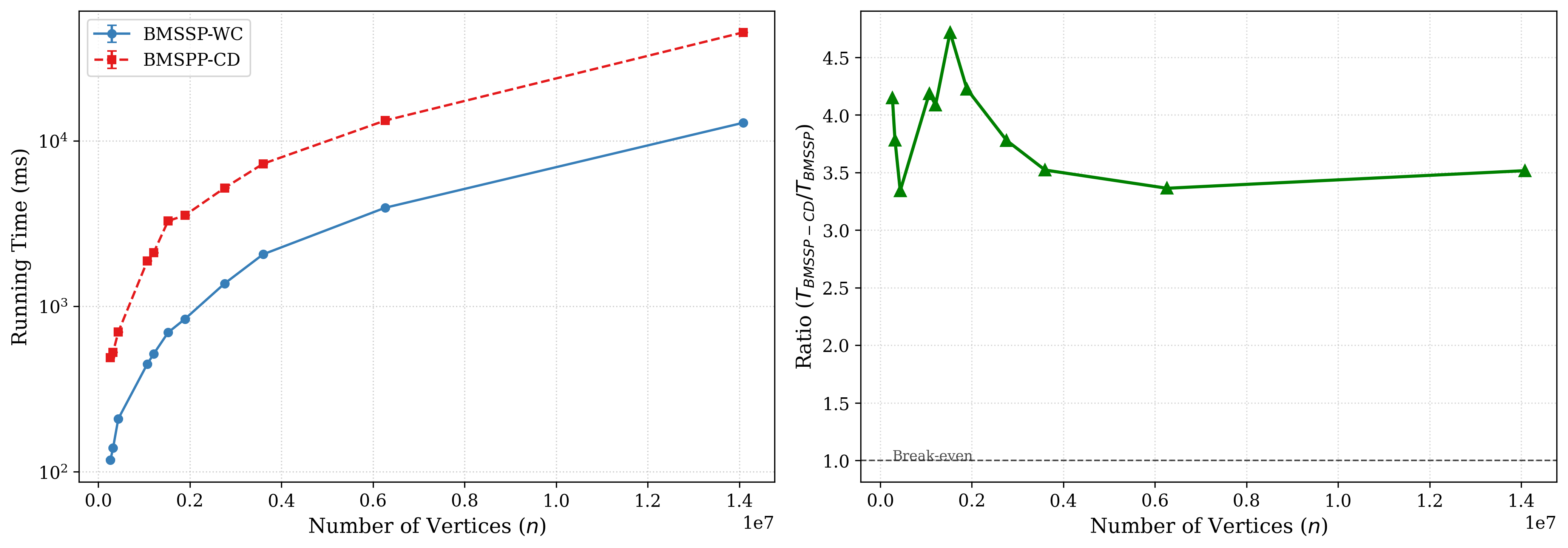}
    \caption{Running times of BMSSP-WC and BMSSP-CD on USA road network instances,
plotted as a function of the number of vertices.
The ratio curve represents the running time of BMSSP-CD divided by that of
BMSSP-WC.}
    \label{fig:ex2_performance_usa}
\end{figure}

For this instance class, BMSSP-CD could not be executed on the largest
graph due to memory limitations.
Consequently, we report results for the remaining 11 instances.
\Cref{fig:ex2_performance_usa} shows the running times of both
algorithms as a function of the number of vertices, together with the
ratio between their execution times.

The results indicate that the constant-degree normalization introduces a
significant overhead in this setting.
As shown in \Cref{tab:ex2_performance_usa}, the performance ratio
stabilizes around $3.5$, with BMSSP-CD consistently slower than BMSSP-WC
across all instances.
For the largest reported instance, BMSSP-WC computes shortest paths in
approximately $12$ seconds, while BMSSP-CD requires approximately
$45$ seconds.

\subsection{H3 Random Graphs}

To evaluate whether degree normalization provides benefits on graphs
without a bounded maximum degree, we also test BMSSP-WC and BMSSP-CD on
random graphs of class H3, in which $3n$ edges are distributed uniformly
at random.
As with the USA road networks, BMSSP-CD could not be executed on the
largest instances due to memory limitations.
As a result, experiments on H3 graphs were conducted up to instances with
$2^{23}$ vertices.

\begin{figure}[!htpb]
    \includegraphics[width=\linewidth]{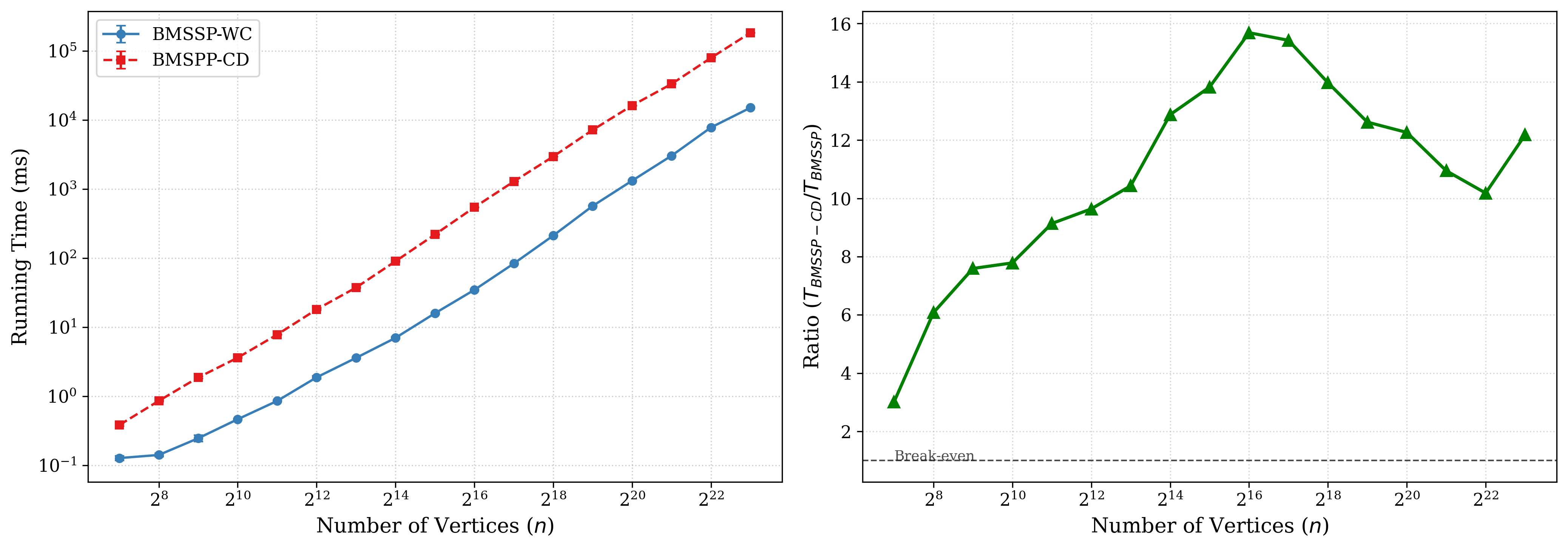}
    \caption{Running times of BMSSP-WC and BMSSP-CD on random graphs H3 instances,
plotted as a function of the number of vertices. The ratio curve represents the running time of BMSSP-CD divided by that of BMSSP-WC.}
    \label{fig:ex2_performance_h3}
\end{figure}

\Cref{fig:ex2_performance_h3} shows the running times of both
algorithms as a function of the number of vertices, together with the
ratio between their execution times.

From a theoretical perspective, one might expect that as the graph size
increases and vertex degrees are unbounded, the degree normalization
could reduce the performance gap between BMSSP-CD and BMSSP-WC.
However, this behavior is not observed experimentally.
Instead, the ratio increases with the number of vertices, and BMSSP-CD
becomes more than $10\times$ slower than BMSSP-WC.

As shown in \Cref{tab:ex2_performance_h3}, the worst-performing
instance occurs at $n = 2^{16}$, where BMSSP-CD is more than $15\times$
slower than BMSSP-WC.
For the largest reported instance, BMSSP-WC computes shortest paths in
approximately $15$ seconds, while BMSSP-CD requires approximately
$183$ seconds.

Overall, these results indicate that, both on sparse road networks with
naturally bounded degrees and on random graphs without bounded maximum
degree, the normalization strategy of BMSSP-CD does not provide practical
performance benefits.
Instead, the additional transformations substantially increase constant
factors, which dominate the overall running time in practice.

\section{Experiment 3 - BMSSP-WC vs. BMSSP-Expected}\label{sec:results3}

In this experiment, we compare the worst-case implementation
(BMSSP-WC) with the expected-time implementation (BMSSP-Expected).
While BMSSP-WC guarantees a worst-case time complexity of
$O(m \log^{2/3} n)$, BMSSP-Expected achieves the same asymptotic bound only in expectation.
The goal of this experiment is to evaluate whether relaxing the
worst-case guarantees leads to practical performance improvements.

\subsection{Random Graphs}

\begin{figure}[!htpb]
    \begin{adjustwidth}{-1.5cm}{-1.5cm}
        \centering
        
        \begin{subfigure}{0.70\linewidth}
            \centering
            \includegraphics[width=\linewidth]{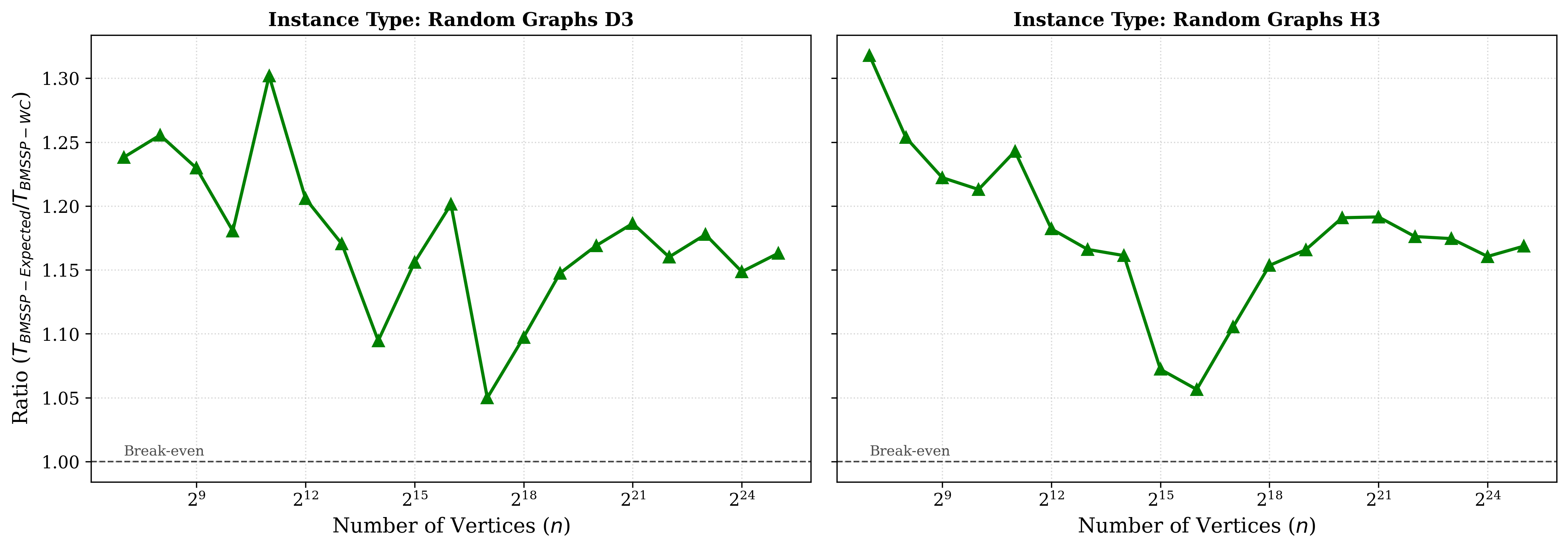}
            \caption{Random sparse graphs of types D3 and H3.}
            \label{fig:ex3_ratio}
        \end{subfigure}

        \vspace{0.1cm}

        \begin{subfigure}{0.70\linewidth}
            \centering
            \includegraphics[width=\linewidth]{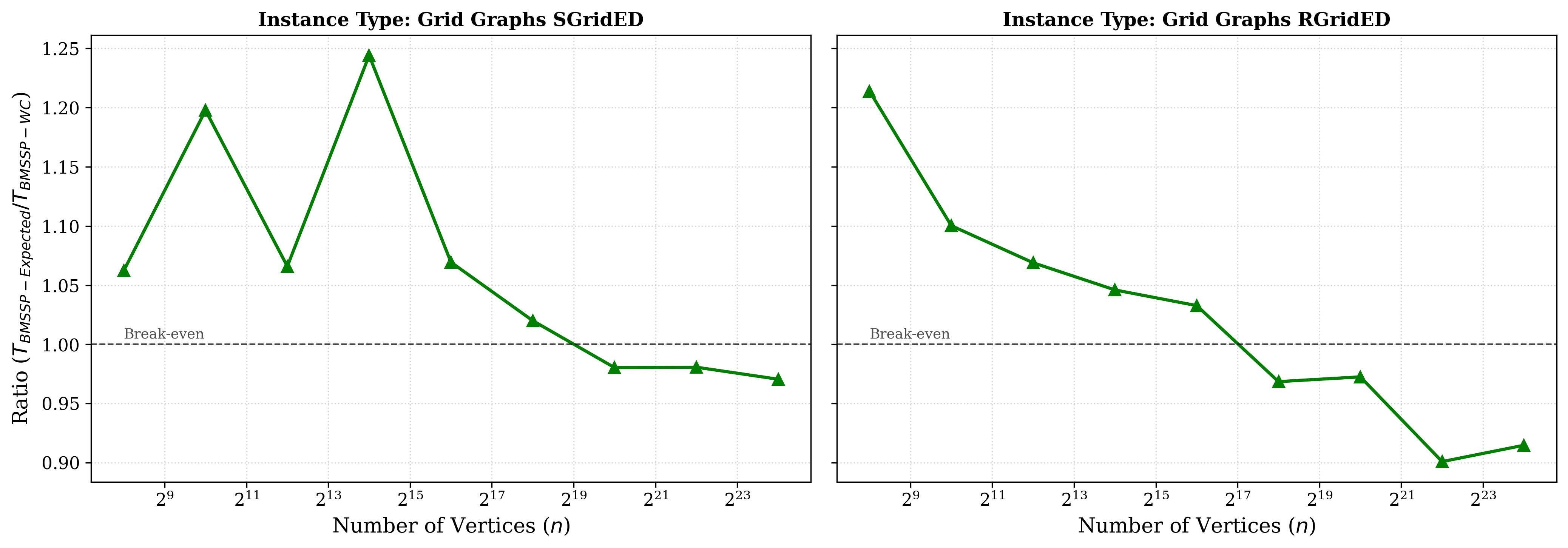}
            \caption{Grid graphs with Euclidean distance (GridED).}
            \label{fig:ex3_ratio_GridED}
        \end{subfigure}

        \vspace{0.1cm}
        
        \begin{subfigure}{0.70\linewidth}
            \centering
            \includegraphics[width=\linewidth]{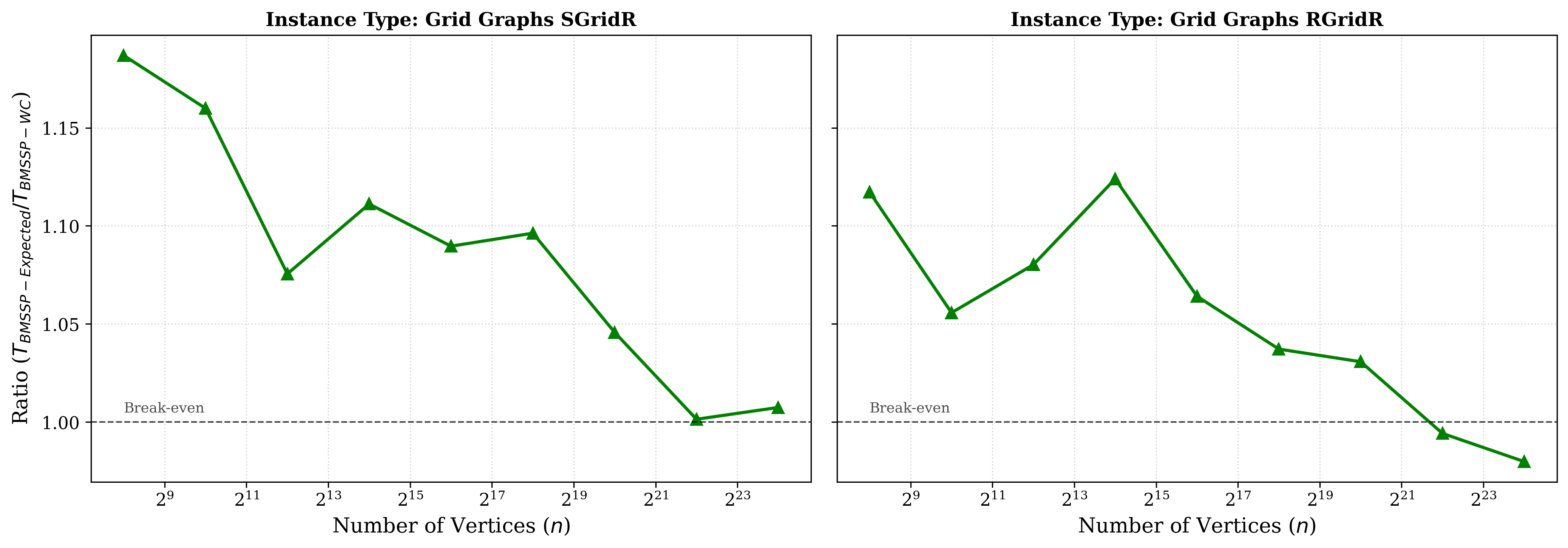}
            \caption{Grid graphs with random distances (GridR).}
            \label{fig:ex3_ratio_GridR}
        \end{subfigure}

        \caption{Observed ratio between the average running times of BMSSP-Expected and BMSSP-WC.}
        \label{fig:combined_all_ratios}
    \end{adjustwidth}
\end{figure}

We first consider random sparse graphs of types D3 and H3.
\Cref{fig:ex3_ratio} shows the ratio between the average running
times of BMSSP-Expected and BMSSP-WC as a function of the number of
vertices.
\Cref{tab:ex3_performance} reports the corresponding numerical
results.

Across all tested instances, BMSSP-Expected is consistently slower than
BMSSP-WC. The performance ratio remains relatively stable, ranging
between $1.05$ and $1.30$ for both D3 and H3. The ratio does not exhibit a clear
 decreasing trend as the number of vertices grows, even for the largest graphs.
This indicates that, although BMSSP-Expected avoids some of the
worst-case-oriented mechanisms of BMSSP-WC, the randomized components do
not translate into practical speedups on these random sparse instances.
Instead, BMSSP-WC remains faster while providing stronger
theoretical guarantees on these graphs.

\begin{figure}[!htpb]
    \centering
    \includegraphics[width=0.5\linewidth]{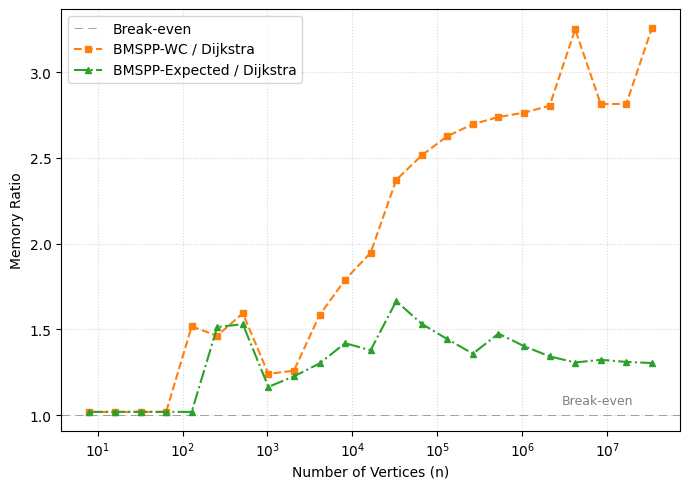}
    \caption{Memory consumption ratios for BMSSP variants vs. Dijkstra on D3 graphs.}
    \label{fig:memory_d3}
\end{figure}

\Cref{fig:memory_d3} shows the maximum memory used by the various implementations on D3 graphs. BMSSP-WC uses nearly $3\times$ more memory than Dijkstra on the largest graphs, while BMSSP-Expected uses approximately $1.5\times$ as much. This highlights the practical impact of the differing space complexities between the implementations.

\subsection{Road Graphs (USA)}

\begin{figure}[!htpb]
    \includegraphics[width=\linewidth]{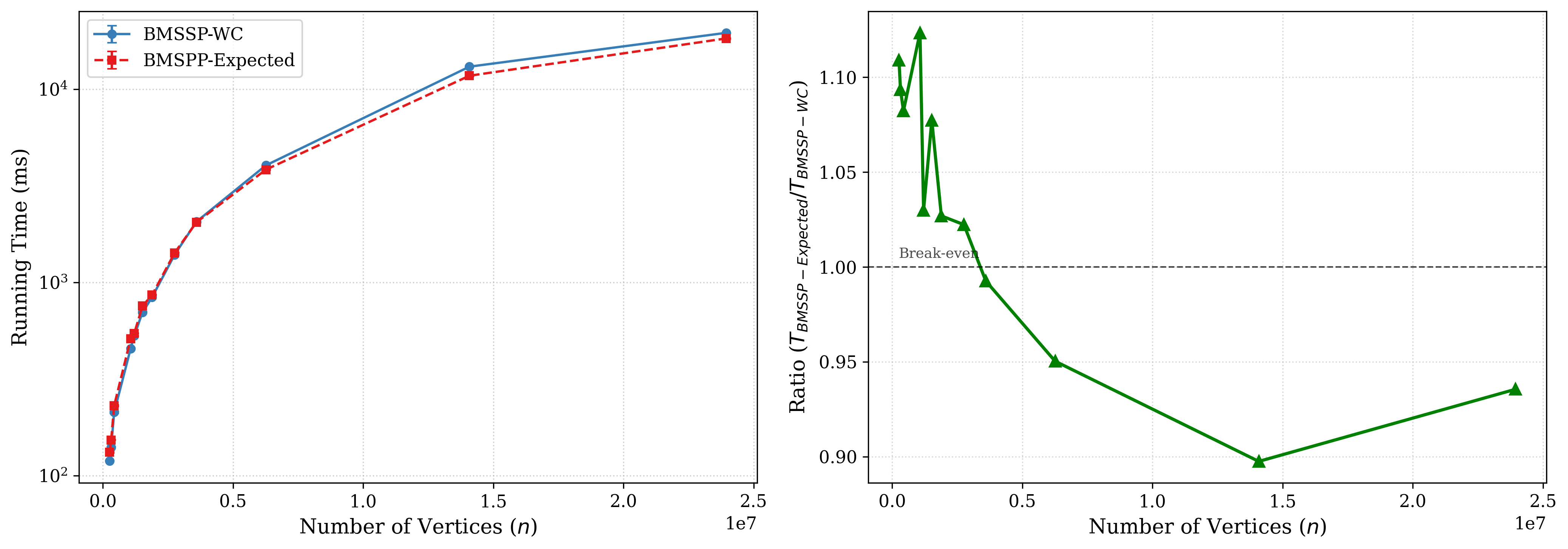}
    \caption{Running times of BMSSP-Expected 
    and BMSSP-WC on USA road
    network instances.}
    \label{fig:ex3_performance_usa}
\end{figure}

We also evaluate BMSSP-WC and BMSSP-Expected on the USA road network
instances.
\Cref{fig:ex3_performance_usa} presents the running times of both
algorithms together with their performance ratio, while
\Cref{tab:ex3_performance_usa} reports the numerical values.

In contrast to the random graphs experiment, the performance gap between the
two implementations is smaller on road networks, and in the largest
instances BMSSP-Expected slightly outperforms BMSSP-WC.
For larger graphs, the ratio drops below $1$, indicating that
BMSSP-Expected becomes faster than BMSSP-WC.

For the largest instance, representing the full USA road network with
approximately $2.3 \times 10^{7}$ vertices and $5.8 \times 10^{7}$ edges,
BMSSP-Expected computes shortest paths in $18.3$ seconds,
compared to $19.6$ seconds for BMSSP-WC; this indicates that BMSSP-Expected is 
around $6\%$ faster for this instance.
Although the observed improvement is modest, it suggests that the
expected-time variant can be competitive on structured real-world graphs
with low average degree, particularly on large road networks such as USA-road-t.USA.

\subsection{Grid Graphs}

\Cref{fig:ex3_ratio_GridED,fig:ex3_ratio_GridR} report the ratio
between the average running times of BMSSP-Expected and BMSSP-WC as a function of
the number of vertices for grid graphs with Euclidean edge weights (GridED) and
random edge weights (GridR), respectively.
\Cref{tab:ex3_performance_GridED,tab:ex3_performance_GridR}
present the corresponding numerical results.

Across both grid models, BMSSP-Expected consistently improves upon the worst-case
variant as the graph size increases.
For small instances, the two algorithms exhibit very similar performance, with
ratios close to~$1$, indicating that the additional overhead of the expected-time
optimizations does not yet amortize.
However, as the number of vertices grows, BMSSP-Expected increasingly outperforms
BMSSP-WC, confirming the effectiveness of the expected-time strategy on grid graphs.

This effect is mainly observed on GridED graphs. For instances with more than $10^6$ vertices, the observed ratio frequently
drops below~$1$, indicating that BMSSP-Expected becomes faster than BMSSP-WC.
For the largest rectangular grids, such as
\texttt{random2048X8192RGridED}, BMSSP-Expected achieves an improvement of nearly
$10\%$ over the worst-case variant.

In the GridR graphs with random integer weights, a similar trend is observed, although the gains are more
moderate and represent around a $2\%$ to $5\%$ improvement for the largest rectangular grid graphs. However, in square grid graphs, BMSSP-WC remains faster for all tested instance sizes.

\section{General Analysis}\label{sec:general}
After running the experiments, it was observed that the classical algorithm of
\citet{dijkstra1959note} achieved superior performance in all evaluated instances.
Although, in theory, the algorithm proposed by \citet{duan2025breaking} admits a better
asymptotic upper bound, its practical implementation is considerably more complex.
It involves multiple recursive calls, additional memory overhead, cache-unfriendly
memory access patterns, and several implementation nuances that introduce a large
constant factor in the running time.

As shown by the experiments, this implementation overhead grows at approximately the
same rate as that of Dijkstra's algorithm, effectively canceling the theoretical
$\log^{1/3} n$ advantage. In practice, this causes the BMSSP-WC implementation to perform
worse than Dijkstra's algorithm, which benefits from a simple, direct structure and
highly optimized implementations, especially when using a \textit{binary heap} stored
in an array.


This behavior is consistently observed not only on random graphs but also on grid
graphs, both with Euclidean and random edge weights.
Despite the strong regularity and low maximum degree of grid graphs, BMSSP-WC does not
outperform Dijkstra in any of the tested instances.
Although the performance gap decreases as the number of vertices increases, no
crossover point is observed even for grids with more than $10^7$ vertices, indicating
that the constant-factor overhead of BMSSP-WC remains dominant on graphs of this magnitude.

As shown in \Cref{ssec:threshold_estimation}, the estimated threshold
$n_0$ for which BMSSP-WC would outperform our Dijkstra implementation is approximately
\bestthresholdpoint{}, in the most optimistic graph type. Such an instance size is computationally infeasible with
current hardware. This result strongly suggests that, for all
practical purposes, Dijkstra's algorithm remains the preferable choice, unless a
significantly more efficient implementation of BMSSP can reduce the constant $c_2$
to a competitive level.

Regarding the second experiment, the degree normalization step makes the use of
BMSSP-CD impractical. The transformation substantially increases the number of
both vertices and edges, leading to a degradation in performance. Instead,
BMSSP-WC and BMSSP-Expected are more useful, as they are built to perform well when the degrees of vertices are small, which is the case for the tested instances.

Finally, compared to BMSSP-WC, BMSSP-Expected shows better performance on large USA road network instances, achieving improvements of up to $20\%$. 
A similar but more moderate advantage of BMSSP-Expected is also observed on grid
graphs, particularly for large and rectangular instances.
In these cases, BMSSP-Expected consistently outperforms BMSSP-WC as the graph size
grows, with improvements of up to $10\%$ on grids with Euclidean edge weights. It also uses less memory than BMSSP-WC as the size graph increases, using $2\times$ less memory on the largest tested graph.

This behavior suggests that the expected-time variant can effectively exploit structural properties of real-world sparse graphs and graphs with small constant degree, making it a promising alternative among the BMSSP variants for urban roads networks.

\section{Conclusion}\label{sec:conclusion}

In this work, we presented the first faithful and fully deterministic implementation
of the algorithm proposed by \citet{duan2025breaking}, which is the first known
comparison-based SSSP algorithm with $o(n \log n)$ worst-case complexity on sparse
graphs. In addition to the worst-case implementation (\textbf{BMSSP-WC}), we implemented
a degree-normalized variant (\textbf{BMSSP-CD}) and an expected-time variant
(\textbf{BMSSP-Expected}), and conducted a large-scale experimental evaluation against
Dijkstra's algorithm using a binary heap.

Our experiments show that, in all tested scenarios, Dijkstra's algorithm was consistently faster, often by a factor of 3 to 4. We predict that for \textbf{BMSSP-WC} to outperform Dijkstra's algorithm, the number of vertices would need to exceed \bestthresholdpoint{}. Regarding algorithmic variants, we observed that degree normalization (\textbf{BMSSP-CD}) has a negative practical impact, while the expected-time variant (\textbf{BMSSP-Expected}) offers performance gains over the worst-case implementation in certain cases and uses close to $2\times$ less memory on the largest tested graph.

Therefore, despite recent theoretical breakthroughs,
Dijkstra's algorithm remains the method of choice for solving the SSSP problem in
practice on large sparse graphs. Our analysis also suggests that only
an implementation of BMSSP with significantly smaller constant factors (if such an
implementation is possible) could make the algorithm competitive with Dijkstra in
practical settings. This work thus establishes a rigorous experimental baseline for
future research on $o(n \log n)$ shortest-path algorithms and highlights the
importance of implementation-level considerations when translating theoretical
improvements into practical performance gains.

\bibliography{references}

\newpage
\begin{appendices}
\counterwithin{table}{section}
\counterwithin{figure}{section}

\FloatBarrier
\section{Experiment 1 Tables}

\begin{table}[!htpb]
    \centering
    \caption{Average running times of Dijkstra's algorithm and BMSSP-WC on random sparse graph instances of types D3 and H3.}
\begin{tabular}{lccccc}
\hline
\multicolumn{1}{c}{\multirow{2}{*}{\textbf{Instance}}} & \multirow{2}{*}{\textbf{$n$}} & \multirow{2}{*}{\textbf{$m$}} & \multicolumn{2}{c}{\textbf{Time (ms)}} & \multirow{2}{*}{\textbf{Ratio}} \\ \cline{4-5}
\multicolumn{1}{c}{} & & & \multicolumn{1}{c}{\cite{dijkstra1959note}} & \cite{duan2025breaking} & \\ \hline
random128D3 & $2^{7}$ & $3\times 2^{7}$ & $0.005 \pm 0.003$ & $0.127 \pm 0.011$ & 25.400 \\
H3random128 & $2^{7}$ & $3\times 2^{7}$ & $0.006 \pm 0.003$ & $0.133 \pm 0.014$ & 22.167 \\ \hline
random256D3 & $2^{8}$ & $3 \times 2^{8}$ & $0.014 \pm 0.004$ & $0.136 \pm 0.008$ & 9.714 \\
H3random256 & $2^{8}$ & $3 \times 2^{8}$ & $0.015 \pm 0.004$ & $0.178 \pm 0.046$ & 11.867 \\ \hline
random512D3 & $2^{9}$ & $3 \times 2^{9}$ & $0.038 \pm 0.004$ & $0.227 \pm 0.012$ & 5.974 \\
H3random512 & $2^{9}$ & $3 \times 2^{9}$ & $0.036 \pm 0.004$ & $0.219 \pm 0.009$ & 6.083 \\ \hline 
random1024D3 & $2^{10}$ & $3 \times 2^{10}$ & $0.087 \pm 0.005$ & $0.504 \pm 0.089$ & 5.793 \\
H3random1024 & $2^{10}$ & $3 \times 2^{10}$ & $0.092 \pm 0.004$ & $0.478 \pm 0.030$ & 5.196 \\ \hline
random2048D3 & $2^{11}$ & $3 \times 2^{11}$ & $0.214 \pm 0.020$ & $0.880 \pm 0.025$ & 4.112 \\
H3random2048 & $2^{11}$ & $3 \times 2^{11}$ & $0.196 \pm 0.005$ & $0.958 \pm 0.065$ & 4.888 \\\hline
random4096D3 & $2^{12}$ & $3 \times 2^{12}$ & $0.432 \pm 0.029$ & $1.818 \pm 0.030$ & 4.208 \\
H3random4096 & $2^{12}$ & $3 \times 2^{12}$ & $0.429 \pm 0.019$ & $1.848 \pm 0.107$ & 4.308 \\\hline
random8192D3 & $2^{13}$ & $3 \times 2^{13}$ & $1.039 \pm 0.045$ & $3.722 \pm 0.091$ & 3.582 \\
H3random8192 & $2^{13}$ & $3 \times 2^{13}$ & $0.962 \pm 0.028$ & $3.586 \pm 0.096$ & 3.728 \\\hline
random16384D3 & $2^{14}$ & $3 \times 2^{14}$ & $2.308 \pm 0.246$ & $6.875 \pm 0.225$ & 2.979 \\
H3random16384 & $2^{14}$ & $3 \times 2^{14}$ & $2.221 \pm 0.050$ & $6.564 \pm 0.046$ & 2.955 \\\hline
random32768D3 & $2^{15}$ & $3 \times 2^{15}$ & $4.675 \pm 0.113$ & $16.356 \pm 0.558$ & 3.499 \\
H3random32768 & $2^{15}$ & $3 \times 2^{15}$ & $4.881 \pm 0.096$ & $15.826 \pm 0.727$ & 3.242 \\\hline
random65536D3 & $2^{16}$ & $3 \times 2^{16}$ & $10.738 \pm 0.199$ & $38.156 \pm 0.801$ & 3.553 \\
H3random65536 & $2^{16}$ & $3 \times 2^{16}$ & $10.482 \pm 0.232$ & $35.381 \pm 0.940$ & 3.375 \\\hline
random131072D3 & $2^{17}$ & $3 \times 2^{17}$ & $29.175 \pm 1.932$ & $81.830 \pm 1.353$ & 2.805 \\
H3random131072 & $2^{17}$ & $3 \times 2^{17}$ & $29.980 \pm 0.243$ & $84.178 \pm 1.065$ & 2.808 \\\hline
random262144D3 & $2^{18}$ & $3 \times 2^{18}$ & $73.481 \pm 0.609$ & $208.062 \pm 0.969$ & 2.832 \\
H3random262144 & $2^{18}$ & $3 \times 2^{18}$ & $74.390 \pm 1.813$ & $213.486 \pm 1.034$ & 2.870 \\\hline
random524288D3 & $2^{19}$ & $3 \times 2^{19}$ & $177.561 \pm 2.665$ & $564.448 \pm 1.586$ & 3.179 \\
H3random524288 & $2^{19}$ & $3 \times 2^{19}$ & $179.838 \pm 1.189$ & $569.684 \pm 2.284$ & 3.168 \\\hline
random1048576D3 & $2^{20}$ & $3 \times 2^{20}$ & $421.216 \pm 1.434$ & $1286.205 \pm 1.306$ & 3.054 \\
H3random1048576 & $2^{20}$ & $3 \times 2^{20}$ & $424.459 \pm 2.710$ & $1317.955 \pm 1.526$ & 3.105 \\\hline
random2097152D3 & $2^{21}$ & $3 \times 2^{21}$ & $939.767 \pm 5.953$ & $2993.859 \pm 9.313$ & 3.186 \\
H3random2097152 & $2^{21}$ & $3 \times 2^{21}$ & $933.016 \pm 6.067$ & $3023.283 \pm 3.785$ & 3.240 \\\hline
random4194304D3 & $2^{22}$ & $3 \times 2^{22}$ & $2022.890 \pm 4.511$ & $7702.933 \pm 44.385$ & 3.808 \\
H3random4194304 & $2^{22}$ & $3 \times 2^{22}$ & $2042.307 \pm 19.507$ & $7718.196 \pm 22.599$ & 3.779 \\\hline
random8388608D3 & $2^{23}$ & $3 \times 2^{23}$ & $4437.921 \pm 6.786$ & $14808.467 \pm 15.531$ & 3.337 \\
H3random8388608 & $2^{23}$ & $3 \times 2^{23}$ & $4519.483 \pm 3.752$ & $14895.695 \pm 18.211$ & 3.296 \\\hline
random16777216D3 & $2^{24}$ & $3 \times 2^{24}$ & $9899.507 \pm 23.569$ & $32367.954 \pm 118.261$ & 3.270 \\
H3random16777216 & $2^{24}$ & $3 \times 2^{24}$ & $10066.866 \pm 47.385$ & $32832.183 \pm 23.318$ & 3.261 \\\hline
random33554432D3 & $2^{25}$ & $3 \times 2^{25}$ & $21812.125 \pm 13.479$ & $82592.131 \pm 48.540$ & 3.787 \\
H3random33554432 & $2^{25}$ & $3 \times 2^{25}$ & $22009.777 \pm 151.378$ & $84702.824 \pm 855.357$ & 3.848 \\
\hline
\end{tabular}
    \label{tab:ex1_performance}
\end{table}

\begin{table}[!htpb]
    \centering
    \caption{Average running times of Dijkstra's algorithm and BMSSP-WC
on USA road network instances.}
\begin{tabular}{lccccc}
\hline
\multicolumn{1}{c}{\multirow{2}{*}{\textbf{Instance}}} & \multirow{2}{*}{\textbf{$n$ (approx.)}} & \multirow{2}{*}{\textbf{$m$ (approx.)}} & \multicolumn{2}{c}{\textbf{Time (ms)}} & \multirow{2}{*}{\textbf{Ratio}} \\ \cline{4-5}
\multicolumn{1}{c}{} & & & \multicolumn{1}{c}{\cite{dijkstra1959note}} & \cite{duan2025breaking} & \\ \hline
USA-road-t.NY & $2.6 \times 10^5 $ & $7.3 \times 10^5$ & $32.406 \pm 0.331$ & $119.361 \pm 0.335$ & 3.683 \\
USA-road-t.BAY & $ 3.2 \times 10^5 $ & $8.0 \times 10^5$  & $38.361 \pm 0.349$ & $140.055 \pm 0.611$ & 3.651 \\
USA-road-t.COL & $4.3 \times 10^5$ & $1.0 \times 10^6$ & $52.109 \pm 0.635$ & $216.572 \pm 1.883$ & 4.156 \\
USA-road-t.FLA & $1.0 \times 10^6$ & $2.7 \times 10^6$ & $128.106 \pm 1.123$ & $455.958 \pm 1.352$ & 3.559 \\
USA-road-t.NW & $1.2 \times 10^6$ & $2.8 \times 10^6$ & $154.594 \pm 1.877$ & $521.545 \pm 1.424$ & 3.374 \\
USA-road-t.NE & $1.5 \times 10^6$ & $3.8 \times 10^6$ & $228.359 \pm 1.810$ & $707.389 \pm 3.497$ & 3.098 \\
USA-road-t.CAL & $1.8 \times 10^6$ & $4.6 \times 10^6$ & $261.384 \pm 2.089$ & $847.049 \pm 2.857$ & 3.241 \\
USA-road-t.LKS & $2.7 \times 10^6$ & $6.8 \times 10^6$ & $381.540 \pm 3.900$ & $1402.351 \pm 3.995$ & 3.676 \\
USA-road-t.E & $3.5 \times 10^6$ & $8.7 \times 10^6$ & $556.746 \pm 6.379$ & $2074.609 \pm 4.951$ & 3.726 \\
USA-road-t.W & $6.2 \times 10^6$ & $1.5 \times 10^7$ & $1022.692 \pm 2.328$ & $3989.016 \pm 9.203$ & 3.901 \\
USA-road-t.CTR & $1.4 \times 10^7$ & $3.4 \times 10^7$ & $3299.535 \pm 9.289$ & $13188.157 \pm 28.296$ & 3.997 \\
USA-road-t.USA & $2.3 \times 10^7$ & $5.8 \times 10^7$ & $4577.503 \pm 8.831$ & $19325.188 \pm 19.417$ & 4.222 \\
\hline
\end{tabular}
    \label{tab:ex1_performance_usa}
\end{table}

\begin{table}[!htpb]
    \centering
        \caption{Average running times of Dijkstra's algorithm and BMSSP-WC on grid graphs with Euclidean distance (GridED) as edge weights.}
\begin{tabular}{lccccc}
\hline
\multicolumn{1}{c}{\multirow{2}{*}{\textbf{Instance}}} & \multirow{2}{*}{\textbf{$n$ (approx.)}} & \multirow{2}{*}{\textbf{$m$ (approx.)}} & \multicolumn{2}{c}{\textbf{Time (ms)}} & \multirow{2}{*}{\textbf{Ratio}} \\ \cline{4-5}
\multicolumn{1}{c}{} & & & \multicolumn{1}{c}{\cite{dijkstra1959note}} & \cite{duan2025breaking} & \\ \hline
random16X16SGridED & $2.5 \times 10^2$ & $1.8 \times 10^3$ & $0.009 \pm 0.003$ & $0.109 \pm 0.011$ & 12.111 \\ 
random8X32RGridED & $2.5 \times 10^2$ & $1.8 \times 10^3$ & $0.006 \pm 0.002$ & $0.165 \pm 0.010$ & 27.500 \\ \hline
random32X32SGridED & $1.0 \times 10^3$ & $7.8 \times 10^3$ & $0.050 \pm 0.005$ & $0.340 \pm 0.023$ & 6.800 \\ 
random16X64RGridED & $1.0 \times 10^3$ & $7.7 \times 10^3$ & $0.037 \pm 0.005$ & $0.669 \pm 0.059$ & 18.081 \\ \hline
random64X64SGridED & $4.0 \times 10^3$ & $3.2 \times 10^4$ & $0.219 \pm 0.008$ & $1.266 \pm 0.114$ & 5.781 \\ 
random32X128RGridED & $4.0 \times 10^3$ & $3.1 \times 10^4$ & $0.180 \pm 0.008$ & $2.806 \pm 0.061$ & 15.589 \\ \hline
random128X128SGridED & $1.6 \times 10^4$ & $1.2 \times 10^5$ & $1.000 \pm 0.057$ & $4.681 \pm 0.085$ & 4.681 \\ 
random64X256RGridED & $1.6 \times 10^4$ & $1.2 \times 10^5$ & $0.974 \pm 0.045$ & $4.641 \pm 0.060$ & 4.765 \\ \hline
random256X256SGridED & $6.5 \times 10^4$ & $5.2 \times 10^5$ & $4.804 \pm 0.352$ & $17.493 \pm 0.186$ & 3.641 \\
random128X512RGridED & $6.5 \times 10^4$ & $5.2 \times 10^5$ & $4.402 \pm 0.197$ & $19.690 \pm 0.175$ & 4.473 \\ \hline
random512X512SGridED & $2.6 \times 10^5$ & $2.0 \times 10^6$ & $21.274 \pm 0.526$ & $80.996 \pm 1.578$ & 3.807 \\ 
random256X1024RGridED & $2.6 \times 10^5$ & $2.0 \times 10^6$ & $21.995 \pm 0.445$ & $98.729 \pm 0.686$ & 4.489 \\ \hline
random1024X1024SGridED &  $1.0 \times 10^6$ & $8.3 \times 10^6$ & $102.265 \pm 1.936$ & $396.114 \pm 0.708$ & 3.873 \\ 
random512X2048RGridED & $1.0 \times 10^6$ & $8.3 \times 10^6$ & $109.942 \pm 2.104$ & $558.391 \pm 0.852$ & 5.079 \\ \hline
random2048X2048SGridED & $4.1 \times 10^6$ & $3.3 \times 10^7$ & $622.311 \pm 1.856$ & $2328.856 \pm 3.873$ & 3.742 \\ 
random1024X4096RGridED & $4.1 \times 10^6$ & $3.3 \times 10^7$ & $646.909 \pm 1.175$ & $3093.970 \pm 2.429$ & 4.783 \\ \hline
random4096X4096SGridED & $1.6 \times 10^7$ & $1.3 \times 10^8$ & $2872.840 \pm 1.642$ & $9860.496 \pm 7.599$ & 3.432 \\ 
random2048X8192RGridED & $1.6 \times 10^7$ & $1.3 \times 10^8$ & $3844.126 \pm 1.976$ & $14431.560 \pm 12.609$ & 3.754 \\
\hline
\end{tabular}
    \label{tab:ex1_performance_GridED}
\end{table}

\begin{table}[!htpb]
    \centering
            \caption{Average running times of Dijkstra's algorithm and BMSSP-WC on grid graphs with random distances (GridR) as edge weights.}
\begin{tabular}{lccccc}
\hline
\multicolumn{1}{c}{\multirow{2}{*}{\textbf{Instance}}} & \multirow{2}{*}{\textbf{$n$ (approx.)}} & \multirow{2}{*}{\textbf{$m$ (approx.)}} & \multicolumn{2}{c}{\textbf{Time (ms)}} & \multirow{2}{*}{\textbf{Ratio}} \\ \cline{4-5}
\multicolumn{1}{c}{} & & & \multicolumn{1}{c}{\cite{dijkstra1959note}} & \cite{duan2025breaking} & \\ \hline
random16X16SGridR & $2.5 \times 10^2$ & $1.8 \times 10^3$ & $0.032 \pm 0.014$ & $0.166 \pm 0.011$ & 5.188 \\
random8X32RGridR & $2.5 \times 10^2$ & $1.8 \times 10^3$ & $0.026 \pm 0.004$ & $0.160 \pm 0.012$ & 6.154 \\ \hline
random32X32SGridR & $1.0 \times 10^3$ & $7.8 \times 10^3$ &$0.112 \pm 0.004$ & $0.622 \pm 0.030$ & 5.554 \\
random16X64RGridR & $1.0 \times 10^3$ & $7.7 \times 10^3$ & $0.128 \pm 0.006$ & $0.550 \pm 0.024$ & 4.297 \\ \hline
random64X64SGridR & $4.0 \times 10^3$ & $3.2 \times 10^4$ & $0.502 \pm 0.025$ & $2.555 \pm 0.051$ & 5.090 \\
random32X128RGridR & $4.0 \times 10^3$ & $3.1 \times 10^4$ & $0.570 \pm 0.024$ & $2.141 \pm 0.033$ & 3.756 \\ \hline
random128X128SGridR & $1.6 \times 10^4$ & $1.2 \times 10^5$ & $2.281 \pm 0.052$ & $8.567 \pm 0.090$ & 3.756 \\
random64X256RGridR & $1.6 \times 10^4$ & $1.2 \times 10^5$ & $2.588 \pm 0.040$ & $8.850 \pm 0.141$ & 3.420 \\ \hline
random256X256SGridR & $6.5 \times 10^4$ & $5.2 \times 10^5$ & $12.731 \pm 0.154$ & $36.239 \pm 0.449$ & 2.847 \\
random128X512RGridR & $6.5 \times 10^4$ & $5.2 \times 10^5$ & $11.337 \pm 0.177$ & $36.273 \pm 0.129$ & 3.200 \\ \hline
random512X512SGridR & $2.6 \times 10^5$ & $2.0 \times 10^6$ & $49.706 \pm 0.546$ & $153.891 \pm 1.423$ & 3.096 \\
random256X1024RGridR & $2.6 \times 10^5$ & $2.0 \times 10^6$ & $62.644 \pm 0.470$ & $165.676 \pm 1.964$ & 2.645 \\ \hline
random1024X1024SGridR &  $1.0 \times 10^6$ & $8.3 \times 10^6$ & $220.996 \pm 1.040$ & $743.351 \pm 0.530$ & 3.364 \\
random512X2048RGridR & $1.0 \times 10^6$ & $8.3 \times 10^6$ & $267.950 \pm 1.616$ & $807.726 \pm 1.281$ & 3.014 \\ \hline
random2048X2048SGridR & $4.1 \times 10^6$ & $3.3 \times 10^7$ & $1072.783 \pm 2.869$ & $4260.375 \pm 2.914$ & 3.971 \\
random1024X4096RGridR & $4.1 \times 10^6$ & $3.3 \times 10^7$ & $1434.521 \pm 3.427$ & $5383.173 \pm 8.856$ & 3.753 \\ \hline
random4096X4096SGridR & $1.6 \times 10^7$ & $1.3 \times 10^8$ & $6126.387 \pm 11.197$ & $19744.810 \pm 5.090$ & 3.223 \\
random2048X8192RGridR & $1.6 \times 10^7$ & $1.3 \times 10^8$ & $7103.698 \pm 25.336$ & $23154.213 \pm 11.196$ & 3.259 \\
\hline
\end{tabular}
     \label{tab:ex1_performance_GridR}
\end{table}

\FloatBarrier
\section{Experiment 2 Tables}

\begin{table}[!htpb]
    \centering
    \caption{Average running times of BMSSP-WC and BMSSP-CD on
USA road network instances.}
    \begin{tabular}{lccccc}
        \hline
        \multicolumn{1}{c}{\multirow{2}{*}{\textbf{Instance}}} & \multirow{2}{*}{\textbf{$n$ (approx.)}} & \multirow{2}{*}{\textbf{$m$ (approx.)}} & \multicolumn{2}{c}{\textbf{Time (ms)}} & \multirow{2}{*}{\textbf{Ratio}} \\ \cline{4-5}
        \multicolumn{1}{c}{} & & & \multicolumn{1}{c}{BMSSP-WC} & BMSSP-CD & \\ \hline
        USA-road-t.NY & $2.6 \times 10^5 $ & $7.3 \times 10^5$ & $118.030 \pm 0.887$ & $489.892 \pm 1.289$ & 4.151 \\
        USA-road-t.BAY & $ 3.2 \times 10^5 $ & $8.0 \times 10^5$  & $139.290 \pm 0.969$ & $526.650 \pm 0.932$ & 3.781 \\
        USA-road-t.COL & $4.3 \times 10^5$ & $1.0 \times 10^6$ & $209.183 \pm 0.704$ & $699.045 \pm 1.644$ & 3.342 \\
        USA-road-t.FLA & $1.0 \times 10^6$ & $2.7 \times 10^6$ &  $448.717 \pm 1.390$ & $1878.650 \pm 3.095$ & 4.187 \\
        USA-road-t.NW & $1.2 \times 10^6$ & $2.8 \times 10^6$ & $515.378 \pm 0.422$ & $2106.777 \pm 5.389$ & 4.088 \\
        USA-road-t.NE & $1.5 \times 10^6$ & $3.8 \times 10^6$ & $695.102 \pm 2.504$ & $3279.082 \pm 4.081$ & 4.717 \\
        USA-road-t.CAL & $1.8 \times 10^6$ & $4.6 \times 10^6$ & $838.086 \pm 9.492$ & $3541.601 \pm 6.947$ & 4.226 \\
        USA-road-t.LKS& $2.7 \times 10^6$ & $6.8 \times 10^6$ & $1373.042 \pm 3.923$ & $5189.072 \pm 9.624$ & 3.779 \\
        USA-road-t.E & $3.5 \times 10^6$ & $8.7 \times 10^6$ & $2060.846 \pm 2.912$ & $7258.218 \pm 3.926$ & 3.522 \\
        USA-road-t.W & $6.2 \times 10^6$ & $1.5 \times 10^7$ & $3936.441 \pm 6.345$ & $13240.946 \pm 21.567$ & 3.364 \\
        USA-road-t.CTR  & $1.4 \times 10^7$ & $3.4 \times 10^7$ & $12832.646 \pm 16.235$ & $45127.778 \pm 13.701$ & 3.517 \\
        \hline
    \end{tabular}
    \label{tab:ex2_performance_usa}
\end{table}

\begin{table}[!htpb]
    \centering
    \caption{Average running times of BMSSP-WC and BMSSP-CD on
random graphs H3.}
    \begin{tabular}{lccccc}
\hline
\multicolumn{1}{c}{\multirow{2}{*}{\textbf{Instance}}} & \multirow{2}{*}{\textbf{$n$}} & \multirow{2}{*}{\textbf{$m$}} & \multicolumn{2}{c}{\textbf{Time (ms)}} & \multirow{2}{*}{\textbf{Ratio}} \\ \cline{4-5}
\multicolumn{1}{c}{} & & & \multicolumn{1}{c}{BMSSP-WC} & BMSSP-CD & \\ \hline
H3random128 & $2^{7}$ & $3\times 2^{7}$ & $0.127 \pm 0.010$ & $0.383 \pm 0.013$ & 3.016 \\
H3random256 & $2^{8}$ & $3\times 2^{8}$ & $0.141 \pm 0.008$ & $0.857 \pm 0.017$ & 6.078 \\
H3random512 & $2^{9}$ & $3\times 2^{9}$ & $0.246 \pm 0.027$ & $1.868 \pm 0.067$ & 7.593 \\
H3random1024 & $2^{10}$ & $3\times 2^{10}$ & $0.464 \pm 0.007$ & $3.612 \pm 0.048$ & 7.784 \\
H3random2048 & $2^{11}$ & $3\times 2^{11}$ & $0.854 \pm 0.017$ & $7.803 \pm 0.068$ & 9.137 \\
H3random4096 & $2^{12}$ & $3\times 2^{12}$ & $1.874 \pm 0.104$ & $18.062 \pm 0.543$ & 9.638 \\
H3random8192 & $2^{13}$ & $3\times 2^{13}$ & $3.587 \pm 0.060$ & $37.423 \pm 0.634$ & 10.433 \\
H3random16384 & $2^{14}$ & $3\times 2^{14}$ & $6.991 \pm 0.174$ & $90.000 \pm 1.680$ & 12.874 \\
H3random32768 & $2^{15}$ & $3\times 2^{15}$ & $15.879 \pm 0.284$ & $219.447 \pm 0.824$ & 13.820 \\
H3random65536 & $2^{16}$ & $3\times 2^{16}$ & $34.660 \pm 0.649$ & $543.691 \pm 1.537$ & 15.686 \\
H3random131072 & $2^{17}$ & $3\times 2^{17}$ & $83.406 \pm 0.295$ & $1286.570 \pm 3.205$ & 15.425 \\
H3random262144 & $2^{18}$ & $3\times 2^{18}$ & $212.024 \pm 0.830$ & $2962.262 \pm 3.572$ & 13.971 \\
H3random524288 & $2^{19}$ & $3\times 2^{19}$ & $569.595 \pm 2.129$ & $7186.526 \pm 2.772$ & 12.617 \\
H3random1048576 & $2^{20}$ & $3\times 2^{20}$ & $1323.220 \pm 1.692$ & $16227.203 \pm 5.317$ & 12.263 \\
H3random2097152 & $2^{21}$ & $3\times 2^{21}$ & $3042.001 \pm 4.565$ & $33300.497 \pm 65.090$ & 10.947 \\
H3random4194304 & $2^{22}$ & $3\times 2^{22}$ & $7810.279 \pm 8.940$ & $79514.480 \pm 101.934$ & 10.181 \\
H3random8388608 & $2^{23}$ & $3\times 2^{23}$ & $15080.451 \pm 10.582$ & $183670.725 \pm 209.920$ & 12.179 \\
\hline
\end{tabular}
\label{tab:ex2_performance_h3}
\end{table}

\FloatBarrier
\section{Experiment 3 Tables}

\begin{table}[!htpb]
    \centering
    \caption{Average running times of BMSSP-Expected and BMSSP-WC on random sparse graph instances of types D3 and H3.}
\begin{tabular}{lccccc}
\hline
\multicolumn{1}{c}{\multirow{2}{*}{\textbf{Instance}}} & \multirow{2}{*}{\textbf{$n$}} & \multirow{2}{*}{\textbf{$m$}} & \multicolumn{2}{c}{\textbf{Time (ms)}} & \multirow{2}{*}{\textbf{Ratio}} \\ \cline{4-5}
\multicolumn{1}{c}{} & & & \multicolumn{1}{c}{BMSSP-WC} & BMSSP-Expected & \\ \hline
random128D3 & $2^{7}$ & $3\times 2^{7}$ & $0.126 \pm 0.010$ & $0.156 \pm 0.016$ & 1.238 \\
H3random128 & $2^{7}$ & $3\times 2^{7}$ & $0.129 \pm 0.009$ & $0.170 \pm 0.022$ & 1.318 \\ \hline
random256D3 & $2^{8}$ & $3 \times 2^{8}$ & $0.137 \pm 0.009$ & $0.172 \pm 0.018$ & 1.255 \\
H3random256 & $2^{8}$ & $3 \times 2^{8}$ & $0.138 \pm 0.009$ & $0.173 \pm 0.017$ & 1.254 \\ \hline
random512D3 & $2^{9}$ & $3 \times 2^{9}$ & $0.209 \pm 0.009$ & $0.257 \pm 0.018$ & 1.230 \\
H3random512 & $2^{9}$ & $3 \times 2^{9}$ & $0.216 \pm 0.009$ & $0.264 \pm 0.016$ & 1.222 \\ \hline 
random1024D3 & $2^{10}$ & $3 \times 2^{10}$ & $0.482 \pm 0.035$ & $0.569 \pm 0.060$ & 1.180 \\
H3random1024 & $2^{10}$ & $3 \times 2^{10}$ & $0.479 \pm 0.008$ & $0.581 \pm 0.033$ & 1.213 \\ \hline
random2048D3 & $2^{11}$ & $3 \times 2^{11}$ & $0.858 \pm 0.026$ & $1.117 \pm 0.127$ & 1.302 \\
H3random2048 & $2^{11}$ & $3 \times 2^{11}$ & $0.869 \pm 0.029$ & $1.080 \pm 0.117$ & 1.243 \\\hline
random4096D3 & $2^{12}$ & $3 \times 2^{12}$ & $1.832 \pm 0.105$ & $2.209 \pm 0.118$ & 1.206 \\
H3random4096 & $2^{12}$ & $3 \times 2^{12}$ & $1.905 \pm 0.111$ & $2.252 \pm 0.132$ & 1.182 \\\hline
random8192D3 & $2^{13}$ & $3 \times 2^{13}$ & $3.763 \pm 0.041$ & $4.404 \pm 0.251$ & 1.170 \\
H3random8192 & $2^{13}$ & $3 \times 2^{13}$ & $3.692 \pm 0.083$ & $4.305 \pm 0.214$ & 1.166 \\\hline
random16384D3 & $2^{14}$ & $3 \times 2^{14}$ & $7.561 \pm 0.710$ & $8.276 \pm 0.355$ & 1.095 \\
H3random16384 & $2^{14}$ & $3 \times 2^{14}$ & $7.134 \pm 0.171$ & $8.284 \pm 0.406$ & 1.161 \\\hline
random32768D3 & $2^{15}$ & $3 \times 2^{15}$ & $15.753 \pm 0.664$ & $18.210 \pm 1.022$ & 1.156 \\
H3random32768 & $2^{15}$ & $3 \times 2^{15}$ & $16.538 \pm 0.351$ & $17.733 \pm 1.130$ & 1.072 \\\hline
random65536D3 & $2^{16}$ & $3 \times 2^{16}$ & $35.969 \pm 0.775$ & $43.211 \pm 1.799$ & 1.201 \\
H3random65536 & $2^{16}$ & $3 \times 2^{16}$ & $36.104 \pm 0.550$ & $38.142 \pm 1.684$ & 1.056 \\\hline
random131072D3 & $2^{17}$ & $3 \times 2^{17}$ & $82.566 \pm 0.246$ & $86.682 \pm 2.868$ & 1.050 \\
H3random131072 & $2^{17}$ & $3 \times 2^{17}$ & $82.402 \pm 0.776$ & $91.082 \pm 2.148$ & 1.105 \\\hline
random262144D3 & $2^{18}$ & $3 \times 2^{18}$ & $212.536 \pm 1.906$ & $233.222 \pm 2.404$ & 1.097 \\
H3random262144 & $2^{18}$ & $3 \times 2^{18}$ & $211.703 \pm 0.751$ & $244.196 \pm 2.115$ & 1.153 \\\hline
random524288D3 & $2^{19}$ & $3 \times 2^{19}$ & $561.520 \pm 1.166$ & $644.347 \pm 7.547$ & 1.148  \\
H3random524288 & $2^{19}$ & $3 \times 2^{19}$ & $577.675 \pm 0.843$ & $673.458 \pm 5.513$ & 1.166 \\\hline
random1048576D3 & $2^{20}$ & $3 \times 2^{20}$ & $1296.598 \pm 1.835$ & $1515.812 \pm 5.960$ & 1.169 \\
H3random1048576 & $2^{20}$ & $3 \times 2^{20}$ & $1320.669 \pm 1.261$ & $1572.659 \pm 8.069$ & 1.191 \\\hline
random2097152D3 & $2^{21}$ & $3 \times 2^{21}$ & $2984.520 \pm 5.767$ & $3540.690 \pm 15.740$ & 1.186 \\
H3random2097152 & $2^{21}$ & $3 \times 2^{21}$ & $3012.154 \pm 3.641$ & $3588.873 \pm 19.718$ & 1.191 \\\hline
random4194304D3 & $2^{22}$ & $3 \times 2^{22}$ & $7622.572 \pm 6.736$ & $8842.806 \pm 53.925$ & 1.160 \\
H3random4194304 & $2^{22}$ & $3 \times 2^{22}$ & $7707.439 \pm 15.096$ & $9064.831 \pm 61.272$ & 1.176 \\\hline
random8388608D3 & $2^{23}$ & $3 \times 2^{23}$ & $14663.760 \pm 36.052$ & $17269.090 \pm 158.983$ & 1.178 \\
H3random8388608 & $2^{23}$ & $3 \times 2^{23}$ & $14915.013 \pm 4.080$ & $17517.535 \pm 136.685$ & 1.174 \\\hline
random16777216D3 & $2^{24}$ & $3 \times 2^{24}$ & $32733.411 \pm 14.179$ & $37598.046 \pm 263.213$ & 1.149 \\
H3random16777216 & $2^{24}$ & $3 \times 2^{24}$ & $32914.541 \pm 20.049$ & $38198.070 \pm 301.511$ & 1.161 \\\hline
random33554432D3 & $2^{25}$ & $3 \times 2^{25}$ & $80508.480 \pm 30.790$ & $93630.580 \pm 429.927$ & 1.163 \\
H3random33554432 & $2^{25}$ & $3 \times 2^{25}$ & $82712.277 \pm 234.263$ & $96659.926 \pm 325.355$ & 1.169 \\
\hline
\end{tabular}
    \label{tab:ex3_performance}
\end{table}

\begin{table}[!htpb]
    \centering
    \caption{Running times of BMSSP-Expected and BMSSP-WC on
USA road network instances.}
    \begin{tabular}{lccccc}
        \hline
   \multicolumn{1}{c}{\multirow{2}{*}{\textbf{Instance}}} & \multirow{2}{*}{\textbf{$n$ (approx.)}} & \multirow{2}{*}{\textbf{$m$ (approx.)}} & \multicolumn{2}{c}{\textbf{Time (ms)}} & \multirow{2}{*}{\textbf{Ratio}} \\ \cline{4-5}
\multicolumn{1}{c}{} & & & \multicolumn{1}{c}{BMSSP-WC} & BMSSP-Expected & \\ \hline
        USA-road-t.NY & $2.6 \times 10^5 $ & $7.3 \times 10^5$ & $119.204 \pm 0.714$ & $132.189 \pm 1.382$ & 1.109 \\
        USA-road-t.BAY & $ 3.2 \times 10^5 $ & $8.0 \times 10^5$  & $139.939 \pm 0.475$ & $153.031 \pm 0.607$ & 1.094 \\
        USA-road-t.COL & $4.3 \times 10^5$ & $1.0 \times 10^6$ & $212.648 \pm 2.056$ & $230.132 \pm 1.977$ & 1.082 \\
        USA-road-t.FLA & $1.0 \times 10^6$ & $2.7 \times 10^6$ & $454.769 \pm 0.896$ & $510.886 \pm 1.705$ & 1.123 \\
        USA-road-t.NW & $1.2 \times 10^6$ & $2.8 \times 10^6$ & $529.171 \pm 1.387$ & $544.993 \pm 1.900$ & 1.030 \\
        USA-road-t.NE & $1.5 \times 10^6$ & $3.8 \times 10^6$ & $700.542 \pm 2.692$ & $754.700 \pm 4.105$ & 1.077 \\
        USA-road-t.CAL & $1.8 \times 10^6$ & $4.6 \times 10^6$ & $837.492 \pm 3.379$ & $860.054 \pm 5.956$ & 1.027 \\
        USA-road-t.LKS& $2.7 \times 10^6$ & $6.8 \times 10^6$ & $1390.777 \pm 3.088$ & $1421.803 \pm 4.308$ & 1.022 \\
        USA-road-t.E & $3.5 \times 10^6$ & $8.7 \times 10^6$ & $2063.483 \pm 4.284$ & $2048.268 \pm 15.613$ & 0.993 \\
        USA-road-t.W & $6.2 \times 10^6$ & $1.5 \times 10^7$ & $4032.778 \pm 4.853$ & $3832.615 \pm 27.774$ & 0.950\\
        USA-road-t.CTR  & $1.4 \times 10^7$ & $3.4 \times 10^7$ & $13077.753 \pm 22.386$ & $11738.670 \pm 74.824$ & 0.898 \\
        USA-road-t.USA & $2.3 \times 10^7$ & $5.8 \times 10^7$ & $19576.321 \pm 7.794$ & $18315.890 \pm 82.088$ & 0.936 \\
        \hline
    \end{tabular}
    \label{tab:ex3_performance_usa}
\end{table}

\begin{table}[!htpb]
    \centering
        \caption{Average running times of BMSSP-WC and BMSSP-Expected on grid graphs with Euclidean distance (GridED) as edge weights.}

\begin{tabular}{lccccc}
\hline
\multicolumn{1}{c}{\multirow{2}{*}{\textbf{Instance}}} & \multirow{2}{*}{\textbf{$n$ (approx.)}} & \multirow{2}{*}{\textbf{$m$ (approx.)}} & \multicolumn{2}{c}{\textbf{Time (ms)}} & \multirow{2}{*}{\textbf{Ratio}} \\ \cline{4-5}
\multicolumn{1}{c}{} & & & \multicolumn{1}{c}{\cite{dijkstra1959note}} & \cite{duan2025breaking} & \\ \hline
random16X16SGridED & $2.5 \times 10^2$ & $1.8 \times 10^3$ & $0.177 \pm 0.022$ & $0.188 \pm 0.015$ & 1.062 \\ 
random8X32RGridED & $2.5 \times 10^2$ & $1.8 \times 10^3$ & $0.145 \pm 0.011$ & $0.176 \pm 0.021$ & 1.214 \\ \hline
random32X32SGridED & $1.0 \times 10^3$ & $7.8 \times 10^3$ & $0.582 \pm 0.032$ & $0.697 \pm 0.061$ & 1.198 \\ 
random16X64RGridED & $1.0 \times 10^3$ & $7.7 \times 10^3$ & $0.499 \pm 0.022$ & $0.549 \pm 0.014$ & 1.100 \\ \hline
random64X64SGridED & $4.0 \times 10^3$ & $3.2 \times 10^4$ & $2.171 \pm 0.093$ & $2.314 \pm 0.090$ & 1.066 \\ 
random32X128RGridED & $4.0 \times 10^3$ & $3.1 \times 10^4$ & $1.988 \pm 0.085$ & $2.125 \pm 0.044$ & 1.069 \\ \hline
random128X128SGridED & $1.6 \times 10^4$ & $1.2 \times 10^5$ & $7.619 \pm 0.044$ & $9.478 \pm 0.280$ & 1.244 \\ 
random64X256RGridED & $1.6 \times 10^4$ & $1.2 \times 10^5$ & $7.550 \pm 0.154$ & $7.897 \pm 0.086$ & 1.046 \\ \hline
random256X256SGridED & $6.5 \times 10^4$ & $5.2 \times 10^5$ & $31.212 \pm 0.082$ & $33.378 \pm 0.702$ & 1.069 \\
random128X512RGridED & $6.5 \times 10^4$ & $5.2 \times 10^5$ & $32.898 \pm 0.346$ & $33.975 \pm 0.983$ & 1.033 \\ \hline
random512X512SGridED & $2.6 \times 10^5$ & $2.0 \times 10^6$ & $146.680 \pm 0.463$ & $149.615 \pm 1.900$ & 1.020 \\ 
random256X1024RGridED & $2.6 \times 10^5$ & $2.0 \times 10^6$ & $144.753 \pm 4.348$ & $140.197 \pm 1.039$ & 0.969 \\ \hline
random1024X1024SGridED &  $1.0 \times 10^6$ & $8.3 \times 10^6$ & $708.683 \pm 0.445$ & $694.776 \pm 6.077$ & 0.980 \\ 
random512X2048RGridED & $1.0 \times 10^6$ & $8.3 \times 10^6$ &  $683.540 \pm 0.633$ & $664.788 \pm 5.521$ & 0.973 \\ \hline
random2048X2048SGridED & $4.1 \times 10^6$ & $3.3 \times 10^7$ & $4030.253 \pm 5.876$ & $3952.511 \pm 107.169$ & 0.981 \\ 
random1024X4096RGridED & $4.1 \times 10^6$ & $3.3 \times 10^7$ & $4117.005 \pm 5.588$ & $3709.408 \pm 9.382$ & 0.901 \\ \hline
random4096X4096SGridED & $1.6 \times 10^7$ & $1.3 \times 10^8$ & $16059.605 \pm 3.952$ & $15587.046 \pm 48.999$ & 0.971 \\ 
random2048X8192RGridED & $1.6 \times 10^7$ & $1.3 \times 10^8$ & $16160.012 \pm 4.526$ & $14783.519 \pm 17.998$ & 0.915 \\
\hline
\end{tabular}
    \label{tab:ex3_performance_GridED}
\end{table}

\begin{table}[!htpb]
    \centering
            \caption{Average running times of BMSSP-WC and BMSSP-Expected on grid graphs with random distances (GridR) as edge weights.}
\begin{tabular}{lccccc}
\hline
\multicolumn{1}{c}{\multirow{2}{*}{\textbf{Instance}}} & \multirow{2}{*}{\textbf{$n$ (approx.)}} & \multirow{2}{*}{\textbf{$m$ (approx.)}} & \multicolumn{2}{c}{\textbf{Time (ms)}} & \multirow{2}{*}{\textbf{Ratio}} \\ \cline{4-5}
\multicolumn{1}{c}{} & & & \multicolumn{1}{c}{BMSSP} & BMSSP-Expected & \\ \hline
random16X16SGridR & $2.5 \times 10^2$ & $1.8 \times 10^3$ & $0.198 \pm 0.012$ & $0.235 \pm 0.042$ & 1.187 \\
random8X32RGridR & $2.5 \times 10^2$ & $1.8 \times 10^3$ & $0.222 \pm 0.011$ & $0.248 \pm 0.013$ & 1.117 \\ \hline
random32X32SGridR & $1.0 \times 10^3$ & $7.8 \times 10^3$ &$0.763 \pm 0.011$ & $0.885 \pm 0.042$ & 1.160 \\
random16X64RGridR & $1.0 \times 10^3$ & $7.7 \times 10^3$ & $0.879 \pm 0.066$ & $0.928 \pm 0.037$ & 1.056 \\ \hline
random64X64SGridR & $4.0 \times 10^3$ & $3.2 \times 10^4$ & $2.939 \pm 0.121$ & $3.161 \pm 0.082$ & 1.076 \\
random32X128RGridR & $4.0 \times 10^3$ & $3.1 \times 10^4$ & $3.179 \pm 0.068$ & $3.434 \pm 0.088$ & 1.080 \\ \hline
random128X128SGridR & $1.6 \times 10^4$ & $1.2 \times 10^5$ & $10.652 \pm 0.173$ & $11.836 \pm 0.310$ & 1.111 \\
random64X256RGridR & $1.6 \times 10^4$ & $1.2 \times 10^5$ & $10.714 \pm 0.093$ & $12.042 \pm 0.089$ & 1.124 \\ \hline
random256X256SGridR & $6.5 \times 10^4$ & $5.2 \times 10^5$ & $42.934 \pm 0.114$ & $46.783 \pm 0.952$ & 1.090 \\
random128X512RGridR & $6.5 \times 10^4$ & $5.2 \times 10^5$ & $45.743 \pm 0.174$ & $48.676 \pm 0.507$ & 1.064 \\ \hline
random512X512SGridR & $2.6 \times 10^5$ & $2.0 \times 10^6$ & $185.904 \pm 0.332$ & $203.802 \pm 4.969$ & 1.096 \\
random256X1024RGridR & $2.6 \times 10^5$ & $2.0 \times 10^6$ & $186.422 \pm 0.729$ & $193.359 \pm 1.340$ & 1.037 \\ \hline
random1024X1024SGridR &  $1.0 \times 10^6$ & $8.3 \times 10^6$ & $852.756 \pm 5.007$ & $891.755 \pm 12.774$ & 1.046  \\
random512X2048RGridR & $1.0 \times 10^6$ & $8.3 \times 10^6$ & $859.999 \pm 1.106$ & $886.459 \pm 3.782$ & 1.031 \\ \hline
random2048X2048SGridR & $4.1 \times 10^6$ & $3.3 \times 10^7$ & $4772.150 \pm 8.222$ & $4779.115 \pm 54.978$ & 1.001 \\
random1024X4096RGridR & $4.1 \times 10^6$ & $3.3 \times 10^7$ & $4602.729 \pm 3.658$ & $4576.182 \pm 13.541$ & 0.994 \\ \hline
random4096X4096SGridR & $1.6 \times 10^7$ & $1.3 \times 10^8$ & $19558.697 \pm 5.573$ & $19704.930 \pm 334.471$ & 1.007 \\
random2048X8192RGridR & $1.6 \times 10^7$ & $1.3 \times 10^8$ & $20760.381 \pm 5.397$ & $20342.425 \pm 52.477$ & 0.980 \\
\hline
\end{tabular}
   \label{tab:ex3_performance_GridR}
\end{table}

\end{appendices}

\end{document}